\documentclass[aps,pre,twocolumn,showpacs,superscriptaddress]{revtex4-1}
\usepackage{amsmath,amssymb}
\usepackage{graphics,graphicx,color}
\usepackage{dcolumn,bm}
\usepackage{tikz}
\usepackage{hyperref}
\usepackage{empheq}
\usepackage{amsfonts}
\usepackage{amsthm}
\usepackage{amssymb}
\usepackage{soul}
\usepackage{url}
\usepackage{caption}
\usepackage{subcaption}

\usepackage{nicematrix}

\graphicspath{ {./figures/} }

\newcommand{\ncd}{\newcommand}
\ncd{\mrm}    {\mathrm}
\ncd{\beq} {\begin{equation}}
  \ncd{\eeq} {\end{equation}}

\newcommand{\matbold}[1]{{\bf{#1}}}

\newcommand{\Am}{\matbold{A}}
\newcommand{\Bm}{\matbold{B}}
\newcommand{\Btm}{\tilde{\matbold{B}}}
\newcommand{\Dm}{\matbold{D}}
\newcommand{\Gm}{\matbold{G}}
\newcommand{\id}{\matbold{I}}
\newcommand{\Mm}{\matbold{M}}
\newcommand{\Pm}{\matbold{P}}
\newcommand{\Qm}{\matbold{Q}}
\newcommand{\Xm}{\matbold{X}}
\newcommand{\Wto}[2]{W_{#2#1}}
\newcommand{\T}{^{\mathrm T}}
\newcommand{\eq}{_{\mathrm{eq}}}
\renewcommand{\u}{\psi}
\newcommand{\uv}{{\bm{\psi}}}
\newcommand{\nr}{N_R}
\newcommand{\gam}{\gamma}

\hypersetup{
  colorlinks,
    linkcolor={red!50!black},
    citecolor={blue!50!black},
    urlcolor={blue!80!black}
  }
  
  \usepackage[bitstream-charter]{mathdesign}
  \DeclareSymbolFont{usualmathcal}{OMS}{cmsy}{m}{n}
\DeclareSymbolFontAlphabet{\mathcal}{usualmathcal}

\begin{document} 




\title{Multifractality and statistical localization in the sparse Barrat--M\'ezard trap model}

\author{Diego Tapias}
\email{diego.tapias@theorie.physik.uni-goettingen.de}
\affiliation{%
Institut f\"ur Theoretische Physik, University of G\"ottingen, Friedrich-Hund-Platz 1, 37077 G\"ottingen, Germany
}%
\author{Peter Sollich}
\email{peter.sollich@uni-goettingen.de}
\affiliation{%
Institut f\"ur Theoretische Physik, University of G\"ottingen, Friedrich-Hund-Platz 1, 37077 G\"ottingen, Germany
}%
\affiliation{Department of Mathematics, King's College London, London WC2R 2LS, UK}

\begin{abstract}
We study within a paradigmatic model for glassy dynamics, the Barrat--M\'ezard trap model, the effect of a nontrivial network structure in the connectivity among traps. 
Sparseness of this network has recently been shown to lead to divergences in the bulk of the spectrum of the associated master operator~\cite{tapias2020entropic,%
tapias2022localization}.
We analyse here specifically the properties of the relaxation modes that contribute to these spectral divergences. We characterize the statistics of the corresponding wavefunctions and demonstrate that they are localized with multifractal properties. The localization patterns are unrelated to the spatial (network) topology, however, and instead fall within the recently introduced class of statistical localization phenomena~\cite{tapias2023multifractality}. To rationalize these results we develop an effective model that successfully explains both the spectral divergences and the power law tails in the wavefunction entries, and provides a clear physical picture of why the localization is statistical rather than spatial.
\end{abstract}
.

\maketitle

\section{Introduction} 

Disordered systems play a central role in modern science~\cite{parisi2023nobel, charbonneau2023spin}. An important family of models within this field is given by random matrix ensembles. These have applications ranging from economics to biology~\cite{potters2020first}, and the corresponding diversity of random matrix models calls for the development of analytical techniques and concepts that help us to understand better e.g.\ the emergence of universal features.

In previous work~\cite{tapias2020entropic, tapias2022localization} we introduced the Sparse Barrat--M\'ezard (BM) trap model. The BM model was originally developed to capture essential properties of glassy dynamics~\cite{barrat1995phase} and falls within the class of trap models~\cite{monthus1996models}. These are described by master operators 
defined on random graphs, with transition rates that satisfy detailed balance but explicitly on the randomly drawn energies (or trap depths) of the nodes. If also the network connectivity is made random and sparse, a remarkable feature that emerges is the existence of divergences within the bulk of the spectral density of the master operator~\cite{tapias2020entropic}. The distribution of the local density of states at the relevant eigenvalues exhibits temperature--dependent power--law tails~\cite{tapias2022localization}. 

Recently, several studies showing a breakdown of Wigner's semicircle law for highly heterogeneous random networks in the high connectivity limit have reported similar phenomenology~\cite{metz2020spectral, silva2022analytic, da2025spectral}, namely the existence of divergences in the bulk of the spectrum, and power--law tails in the distribution of the local density of states, with the tail exponents controlled in this case by heterogeneity in the node degrees. In an analysis of the wavefunction statistics of the modes that contribute to the spectral divergence~\cite{tapias2023multifractality} we found that those modes are localized, with multifractal properties. We dubbed the corresponding localization mechanism ``statistical localization'' because, in contrast to Anderson localization, there is no unique localization centre on the network around which the amplitude of the wavefunction is concentrated: instead, there are multiple centres whose location is determined by the statistical properties of the corresponding nodes.

In this work, we first perform a multifractal analysis of the sparse BM trap model to understand the wavefunction statistics of the modes that contribute to divergences in the spectral bulk. The outcome is that the distribution of (squared) wavefunction entries has power law tails towards large and small values, with temperature--dependent exponents that match those for the local density of states. In the second and main part of the paper, we then construct an effective model that allows us to understand the qualitative physics we observe. This effective model provides a clear physical intuition for the origin of localization in the relaxation modes concerned, and shows that the mechanism at work is indeed statistical localization.

\section{Sparse Barrat--M\'ezard trap model}
\label{sec:bm}

In this section, we recall the main features of the sparse Barrat--M\'ezard (BM) model  that will be needed for the subsequent analysis. For details we refer the reader to Refs.~\cite{tapias2020entropic, tapias2022localization}.

The sparse BM model describes stochastic dynamics on a network (or graph) of $N$ nodes. Physically, these nodes represent local minima in the rough energy landscape of a glassy particle system. The connectivity of the network is defined by an adjacency matrix $\Am$, with $A_{ij}=A_{ji}=1$ if there is an edge between nodes $i$ and $j$ and $A_{ij}=0$ otherwise (including $A_{ii}=0$). The degree of each node is the number $k_i=\sum_j A_{ij}$ of neighbours it has on the network; we denote by $c = \frac{1}{N}\sum_i k_i$ the average degree.

The dynamics is defined by the rate for stochastic transitions between any two nodes $i\neq j$:
\begin{align}
  M_{ji} = \frac{A_{ji}}{c} \frac{1}{1 + \exp(-\beta(E_j - E_i))}
  \label{mast}
\end{align}
The factor $A_{ij}$ means transitions are possibly only along network edges. The fraction gives a Glauber dependence on the energy of each trap, which we specify by the {\em trap depth} $E_i$, i.e.\ its energy counted downwards from some reference value; $\beta=1/T$ is the inverse temperature. Transitions to significantly deeper traps ($E_j-E_i \gtrsim T$) thus happen at constant rate $1/c$, while those to shallower traps ($E_i-E_j \gtrsim T$) are suppressed by an Arrhenius factor. 

The sparse BM model as defined above contains disorder in the trap depths $E_i$ and in the network structure $A_{ij}$. The standard assumption for the trap depths is that they are drawn independently for each trap, from an exponential distribution $\rho_E(E) = \frac{1}{T_{\rm{g}}} {\rm{e}}^{-E/{T_{\rm{g}}}}$ (for $E>0$). 
The energy scale $T_{\rm{g}}$ can be shown to set the glass transition temperature of the model~\cite{barrat1995phase}; we work throughout with energy units so that $T_{\rm{g}}=1$. As regards the network structure, for  concreteness and as in previous studies~\cite{tapias2020entropic, tapias2022localization} we will focus on random regular graphs, i.e.\ a network ensemble that assigns uniform probability across all adjacency matrices for which all nodes have the same degree $k_i=c$. For numerical calculations we generally use 
connectivity $c = 3$, as small $c$--values make network effects more pronounced and so easier to detect.


The time evolution can be described in terms of the probabilities $p_i(t)$ of being in trap $i$ at time $t$. The vector ${\bm{p}}(t) = (p_1(t), \ldots, p_N(t))$ obeys the master equation 
\begin{align}
  \frac{d {\bm{p}} (t)}{dt} = \Mm {\bm{p}} (t)
\end{align}
Here the master operator $\Mm$ has as off-diagonal elements the transition rates~\eqref{mast}, while its diagonal elements
\begin{equation}
M_{ii} = - \sum_{j \neq i} M_{ji}
\label{M_diag}
\end{equation}
are the (negative) exit rates from each node. The eigenvalues of $\Mm$ are then the relaxation rates of the model and the corresponding eigenvectors are the relaxation modes.

To proceed with analytical techniques for random matrices it is useful to switch to a symmetrized form of the master operator $\Mm$. The symmetrization of  $\Mm$ rests on the fact that the dynamics defined by~\eqref{mast} obeys detailed balance with respect to the (Boltzmann) steady state distribution. Defining $\Pm\eq$ as a diagonal matrix containing the Boltzmann probabilities, $(\Pm\eq)_{ii} =Z^{-1}{\rm{e}}^{\beta E_i}$ with $Z$ the partition function, detailed balance implies that the matrix
\begin{equation}
{\Mm}^s = \Pm\eq^{-1/2} {\Mm}  \Pm\eq^{1/2}
\label{M_symm}
\end{equation}  
is symmetric.
This symmetrization does not change the spectrum of the operator, i.e.\ $\Mm^s$ and $\Mm$ have the same eigenvalues. The eigenvectors are also directly related: $\uv$ is an eigenvector of $\Mm^s$ if and only if $\Pm\eq^{1/2}\uv$ is a right eigenvector of $\Mm$.

\subsection{Spectral density and divergences in the bulk}
\label{divergences_recap}

\begin{figure}
  \centering
  \includegraphics[width=0.4\textwidth]{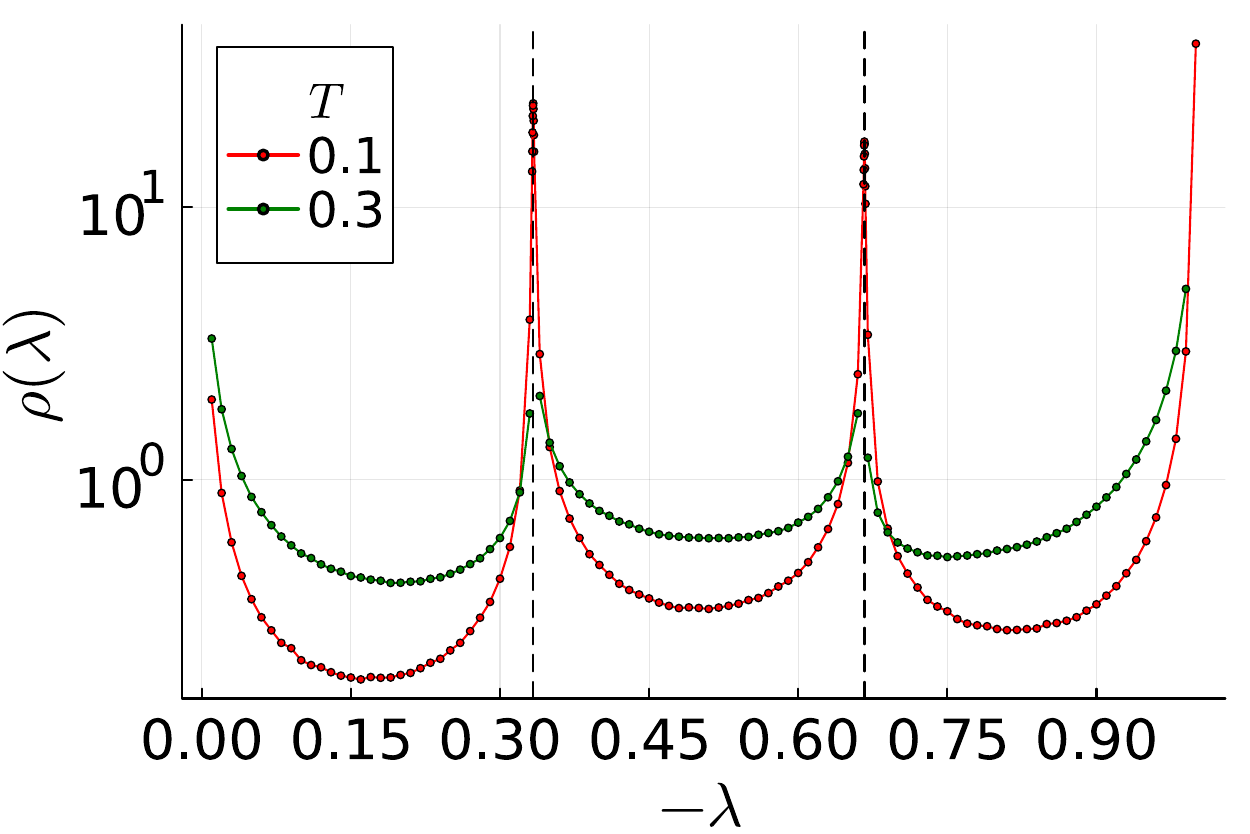}
  \caption{Spectral density $\rho(\lambda)$ of the sparse BM trap model on a random regular graph with connectivity $c=3$ for two different temperatures, evaluated using population dynamics (population size $10^5$) on a $\lambda$--grid with $\Delta \lambda = 0.01$ and with a regularizer $\epsilon = 10^{-3}$. Note the peaks at values of $-\lambda$ that are integer multiples of $1/c=1/3$, where $\rho(\lambda)$ diverges for low enough $T$.}
  \label{spectrum}
\end{figure}

The spectral density $\rho(\lambda)$ is the distribution of the eigenvalues of the master operator in the limit $N \to \infty$. In Figure~\ref{spectrum} we show this spectral density, obtained using the cavity method with population dynamics, for two different temperatures. (For details of the cavity method and its implementation for the sparse BM model, see Ref.~\cite{tapias2020entropic}.) As the figure demonstrates, the spectrum depends strongly on the temperature $T$ as the key control parameter. In particular, for low temperatures, divergences appear at  $\lambda =  \{0, -1/c, \ldots 1\}$, i.e.\ at integer multiples of $-1/c$.

The divergence at $\lambda = 0$ is in fact the same as the one appearing in the sparse Bouchaud trap model~\cite{margiotta2018spectral, riccardothesis} and is characteristic of thermally activated relaxation behavior on long timescales. It can be shown analytically that the divergence around this point is of the form $\rho(\lambda) \sim |\lambda|^{T-1}$~\cite{tapias2020entropic} . Looking at the divergences also at other $\lambda$, a way to understand these is to consider the $T \to 0$ limit~\cite{barrat1995phase,tapias2020entropic}. In this regime, the Glauber factors in the transition rates~\eqref{mast} become Heaviside functions so only transitions to deeper traps occur, at constant rate $1/c$. If the nodes are numbered so that the trap depths are in ascending order $E_1<E_2<\ldots<E_N$, the master operator $\Mm$ is therefore lower triangular and its eigenvalues are the negative exit rates $M_{ii}$. These exit rates are of the form $-n/c$ where $n$ is the number of neighbours of node $i$ that are deeper traps, e.g.\ $n=0$ for a local minimum (node surrounded by shallower traps), $n=1$ for a saddle of index 1 (node with one ``downhill'' exit direction) etc. The spectrum therefore consists of a sum of delta functions at $\lambda=0,-1/c,\ldots,1$.
For finite temperatures these
delta functions broaden to power law  divergences and at higher $T$ to smooth maxima. Eventually also these washed out; for $T\to \infty$ the master operator becomes a rescaled Laplacian and the spectrum is given by the Kesten--McKay law~\cite{mckay1981expected}.

In contrast with the limit $\lambda \to 0$, we so far do not have a theory that predicts the specific form of the spectral divergences around nonzero $\lambda$. In Ref.~\cite{tapias2020entropic} the somewhat simpler spectrum of exit rates for finite temperatures was studied and shown to exhibit similar power--law divergences, without however leading to a prediction for the power--law exponents (for non--trivial connectivity $c > 2$). For the spectral density $\rho(\lambda)$ of relaxation rates, numerical data around $\lambda = -1/c$ suggested the scaling
\begin{align}
  \rho(\lambda) \sim |\lambda + 1/c|^{2T - 1}
  \label{specd}
\end{align}
and thus a divergence only for temperatures $T < 1/2$. One of our aims in this paper is to understand the emergence of this power law scaling and the reason for the qualitative change at a temperature ($T=1/2$) well below the glass transition at $T=1$. 

We note in passing that the temperature $T = 1/2$ marks a change in behaviour also in the mean-field BM model~\cite{barrat1995phase}, though for a different physical reason.
In the mean-field model, $T = 1/2$ is the temperature above which the relaxation dynamics is driven by thermal activation, while for $T < 1/2$ it is predominantly entropic~\cite{bertin2003cross}. In contrast, in the sparse model, $T = 1/2$ signals the threshold for a divergence in the spectral density $\rho(\lambda)$ around $\lambda = -1/c$.


\subsection{Resolvent statistics around $\lambda = -1/c$}

To understand the nature of the relaxation modes at the spectral divergences, we performed in Ref.~\cite{tapias2022localization} 
an analysis of the resolvent statistics for $\lambda = -1/c$. The resolvent is defined as the complex matrix
\begin{align}
  \Gm(\lambda - i \epsilon) = \left((\lambda - i \epsilon) {\id} - {\Mm}^s\right)^{-1}
\end{align}
with $\id$ the identity matrix and $\epsilon$ a small regularizer. The imaginary part of the diagonal resolvent entries, $y_i = {\rm{Im}}(G_{ii})$, is known as Local Density of States (LDoS). Its average over all nodes is proportional to the spectral density, while its distribution provides information about the localization of the corresponding modes~\cite{metz2010localization, susca2021cavity}. For a given matrix $\Mm^s$, the relation between the LDoS at node $i$ and the corresponding eigensystem is
\begin{align}
  y_i = \sum_{\mu = 1}^N |\psi^{(\mu)}_i|^2\,\frac{\epsilon }{(\lambda - \lambda_\mu)^2 + \epsilon^2}
 \label{ldos_def} 
\end{align}
Here the $\lambda_\mu$ are the eigenvalues of $\Mm^s$ and the $\{ \bm{\psi}^{(\mu)}\}$ the corresponding eigenvectors, with components $\psi^{(\mu)}_i$. 

\begin{figure}
  \centering 
  \includegraphics[width=0.4\textwidth]{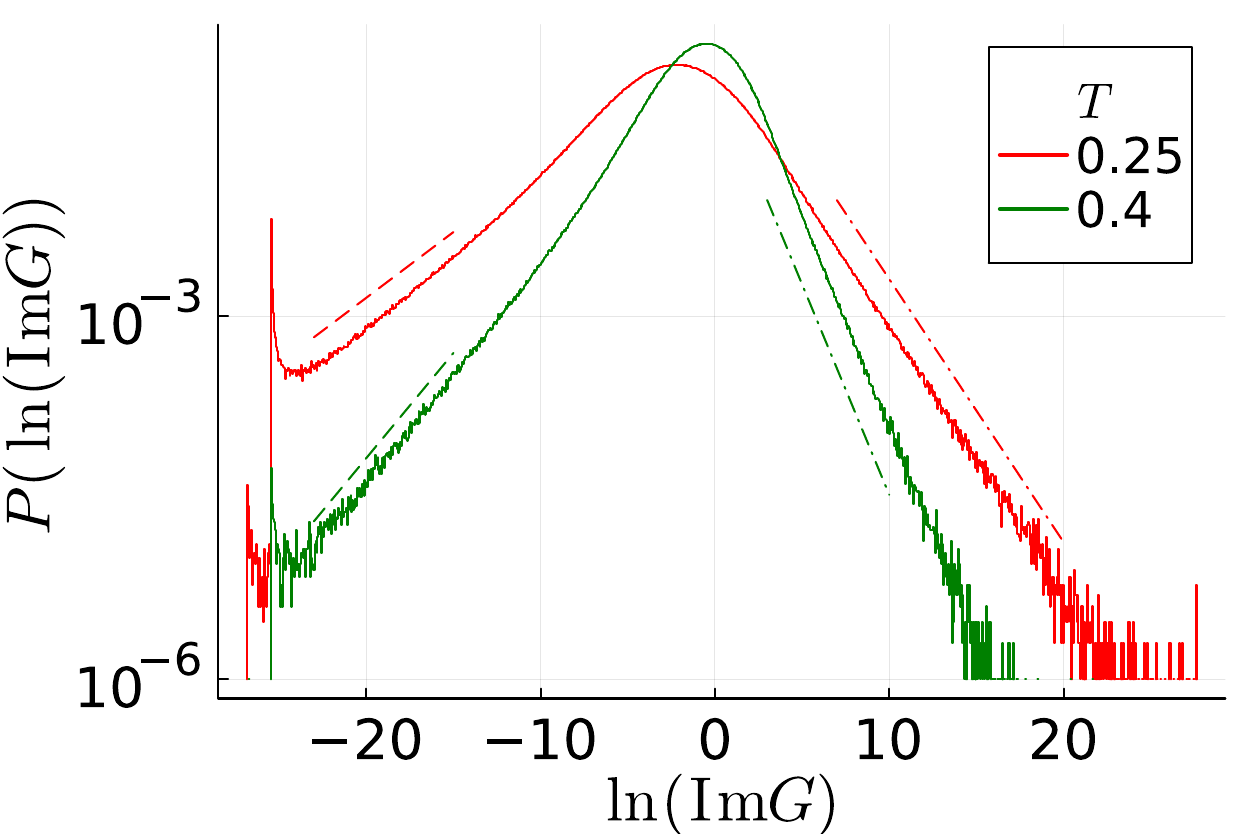}
  \caption{Distribution of the (logarithm of the) local density of states at $\lambda = -1/3$ for two different temperatures, obtained using population dynamics (population size $10^5$, regularizer $\epsilon = 10^{-12}$, connectivity $c=3$). The dash--dotted lines show the power--law scaling~\eqref{resolvent}.}
  \label{resolvent_fig}
\end{figure}

The numerical results for the LDoS distribution at $\lambda=-1/c$ in Ref.~\cite{tapias2022localization} (see also Fig.~\ref{resolvent_fig}) show that this exhibits a fat (power--law) tail towards large values with a temperature dependent exponent, specifically
\begin{align}
  P^{\rm{right}}_{-1/3}(y) \sim  y^{-(2T +1)}
  \label{resolvent}
\end{align}
for $y \gg 1$. 
Inspection of the distribution of the $y_i$ across individual numerically generated networks showed that the largest values occur on nodes surrounded by two shallower traps and one deeper trap, i.e.\ on saddles of index 1 as defined above. 
This makes sense physically because for $T \to 0$, as explained above, it is precisely these index--1 saddles that contribute to the delta function in the spectral density at the eigenvalue being considered, i.e.\ $\lambda=-1/c$. 
This insight is also the starting point for the effective model that we will construct in the second part of this manuscript. 

We observe in Fig.~\ref{resolvent_fig} that the LDoS-distribution exhibits power-law behavior also for small values, again with a temperature-dependent exponent. 
The numerical results (see Fig.~\ref{resolvent_fig}) specifically suggest the scaling
\begin{align}
  P^{\rm left}_{-1/3}(y) \sim y^{T - 1}
  \label{left tail}
\end{align}
This \emph{ansatz} is further motivated by the statistics of the wavefunction amplitudes, as discussed in the following section.

We comment briefly on the contrast to Anderson localization on random regular graphs (see, for instance, Refs.~\cite{mirlin1994distribution, kravtsov2018non, tikhonov2019critical}). There, the LDoS distribution in the localized phase exhibits a right tail with a universal exponent of $-3/2$ while for small values it is essentially exponential, and the typical value of the LDoS scales with $\epsilon$.
Here, on the other hand,
the LDoS distribution at $\lambda = -1/c$ is essentially $\epsilon$-independent 
(see also the discussion in Ref.~\cite{tapias2022localization})
and exhibits two-sided, temperature-dependent power-law behavior. Thus, the relaxation modes associated with this spectral divergence, and by extension the ones at other multiples of $-1/c$, must possess qualitatively distinct localization properties, which we aim to analyze below~\footnote{
For values of $\lambda$ away from the spectral divergences at multiples of $-1/c$, the situation is different in that the numerical evidence points to an Anderson-like delocalization--localization transition at sufficiently low temperatures~\cite{tapias2022localization}.}.


\section{Wavefunction statistics}

The results from~\cite{tapias2022localization} for the LDoS distribution at the spectral divergences (specifically, $\lambda=-1/c$), summarized above, do not immediately lend themselves to an interpretation in terms of the properties of individual relaxation modes. This is because the LDoS~\eqref{ldos_def} effectively mixes a large number of eigenvectors, of order $N\epsilon\rho(\lambda)$. We therefore first analyse the statistics of individual relaxation modes (``wavefunctions'') $\uv$ and extract their multifractal properties using analysis tools from random matrix theory.

\begin{figure}
        \begin{subfigure}[b]{0.48\textwidth}
            \centering
            \includegraphics[width=\textwidth]{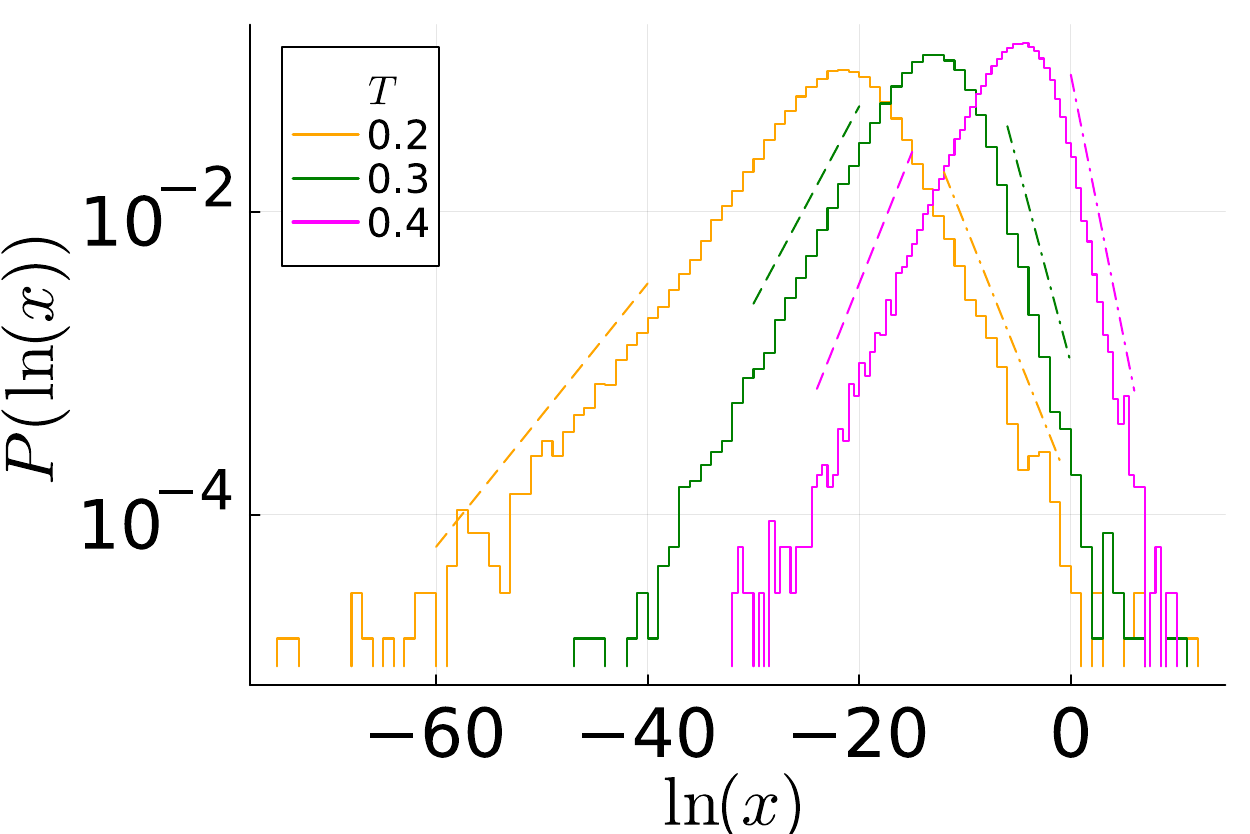}
        \end{subfigure}
        \begin{subfigure}[b]{0.48\textwidth}
            \centering
            \includegraphics[width=\textwidth]{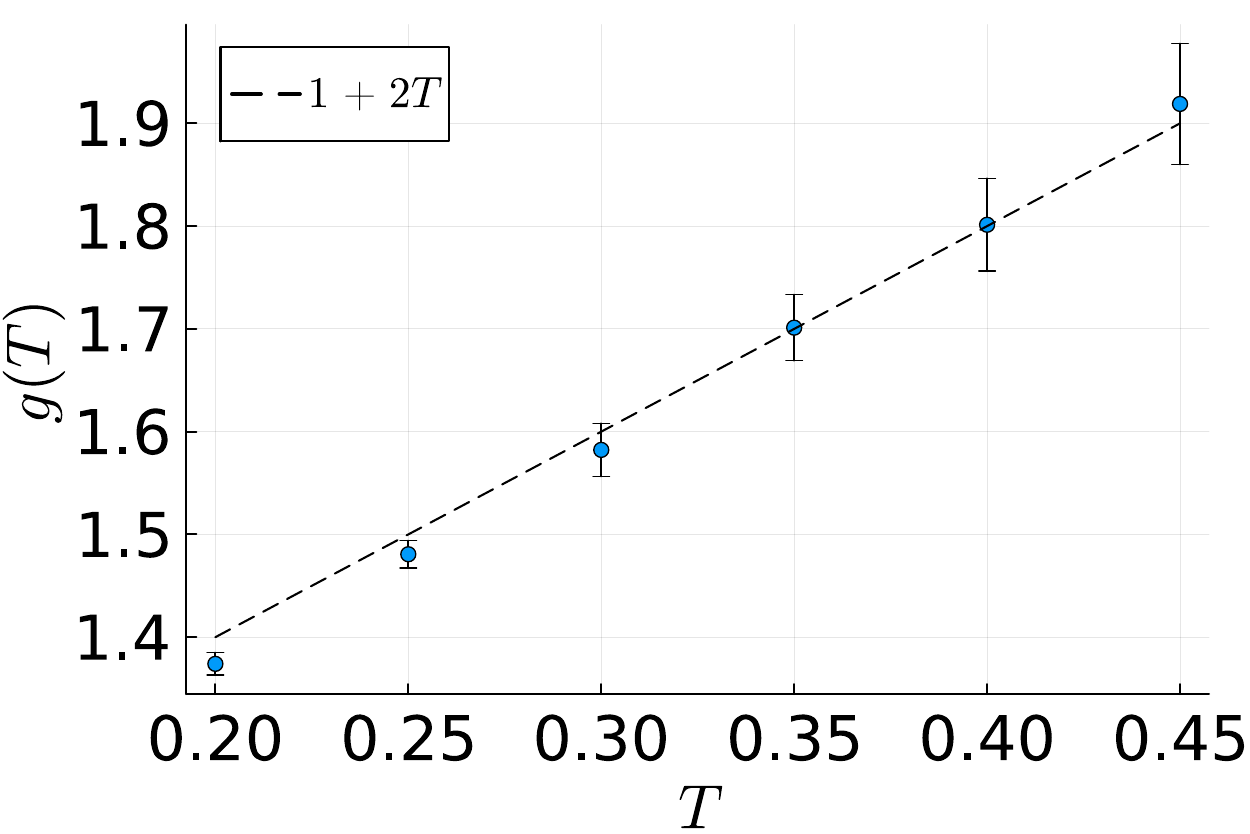}
        \end{subfigure}
        \caption{(a) Distribution of the (logarithm of the) wavefunction amplitudes for three different eigenvectors of the symmetrized master operator ${\Mm}^s$, each corresponding to a separate realization of ${\Mm}^s$ with system size $N = 2^{16}$ and generated at different temperatures. For each sample, the eigenvector shown corresponds to the eigenvalue closest to $\lambda = -1/3$. Dash--dotted lines show the scaling~\eqref{px} for large $x$, while dashed lines correspond to the scaling~\eqref{pxleft} of the left tail.  (b) Numerical estimates of the exponent \( g(T) \) of the right power-law tail \( P_{\rm{right}}(x) \sim x^{-g(T)} \) are shown for various temperatures with \( T < 1/2 \). For each eigenvector (we analyse the 50 eigenvectors with eigenvalues closest to \( \lambda = -1/3 \) in each instance of \( \Mm^s \)), we compute the distribution of \( x \) and estimate the exponent \( g(T) \) by fitting a power-law to the range \( x > x^* \), with \( x^* = 20\, x^{\rm{typ}} \), well above the mode of the distribution. The final results are averaged over all eigenvectors and 10 disorder realizations. The data are consistent with the prediction \( g(T) = 1 + 2T \); error bars indicate the variation across instances.    }
        \label{tail}
      \end{figure}

Numerically, we generate instances of the master operator for finite $N$ by sampling a network structure from the random regular graph ensemble (with $c=3$), drawing the trap depths $E_i$ from $\rho_E(E)$ and calculating $\Mm$ using~(\ref{mast},\ref{M_diag}). After symmetrization~\eqref{M_symm}
we 
diagonalize $\Mm^s$ with the Lanczos algorithm to extract the eigenvectors with eigenvalues closest to $-1/3$. In Fig.~\ref{tail}(a) we show the resulting distribution of the (scaled) squared wavefunction amplitudes $x_i = N|\psi_i|^2$, aggregated from a number of independent instances of $\Mm^s$ and for three different values of $T$. The distributions exhibit two power law tails, each with $T$ dependent exponents. The exponent of the right tail, in particular, matches the one of the LDoS distribution~\eqref{resolvent} as shown in Fig.~\ref{tail}(b).

\begin{figure}
  \centering
  \includegraphics[scale=0.4]{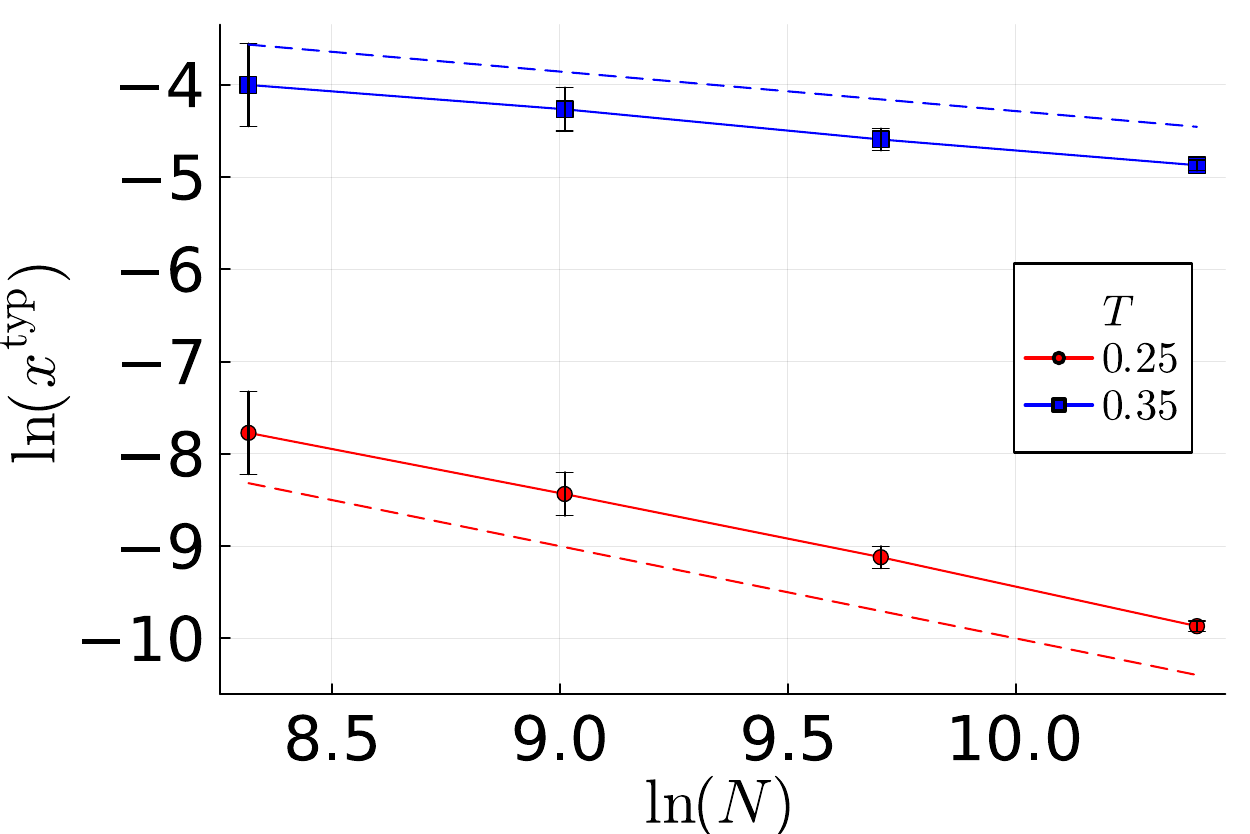}
 \caption{Scaling of the typical value \( x^{\rm{typ}} \) with system size \( N \) for two different temperatures. The theoretical prediction \( x^{\rm{typ}} \sim N^{1 - 1/2T} \) from equation~\eqref{typ_value} is shown as dashed lines. For each realization of \( \Mm^s \), we compute \( x^{\rm{typ}} \) separately for each of the 50 eigenvectors with eigenvalues closest to \( \lambda = -1/3 \), and average the results. Error bars reflect the standard deviation over \( 2^{21}/N \) disorder instances.}
  \label{typ_scaling}
\end{figure}

To further characterize the distribution $P(x)$ we need to know the $N$-dependence of the  typical value $x^{\rm{typ}}$, defined as $x^{\rm{typ}}=\exp(\langle\ln x\rangle)$. In Fig.~\ref{tail}(a) this corresponds roughly to the maximum of $P(\ln x)$, and for $x>x^{\rm typ}$ we have to a reasonable approximation
       \begin{align}
            P_{\rm{right}}(x) \approx  A_N x^{- (2T+1)}  
        \label{px}
          \end{align}
with some $N$-dependent normalization constant $A_N$. 
The  normalization integral $\int dx\,P(x)$ is then dominated by the region $x>x^{\rm typ}$ and evaluates to $\sim A_N (x^{\rm typ})^{-2T}$, showing that $A_N\sim  (x^{\rm typ})^{2T}$. The normalization of $\uv^2$ requires further that $1=\int dx\,x\,P(x)$. If we assume that the scaling~\eqref{px} holds up to the maximum value $x=N$, this integral is dominated by its upper end and scales as $A_N N^{1-2T}$. Altogether we have the scalings
\begin{equation}
A_N \sim N^{2T-1}, \qquad 
            x^{\rm{typ}} \sim N^{1 - 1/(2T)}
            \label{typ_value}
\end{equation}
In Fig.~\ref{typ_scaling} we confirm this scaling of the typical value numerically. Equation~\eqref{px} together with~\eqref{typ_value} fully describes the right tail of the distribution $P(x)$ and is enough to obtain the set of multifractal exponents for positive $q$ (see eq.~\eqref{moment_scaling} below). 

We consider next the left tail of $P(x)$. As shown by the dashed lines in  
Fig.~\ref{tail}(a), this also has a power law form. 
Numerically we find that the exponent is consistent with  $P(\ln x) \sim (\ln x)\T$ or equivalently $P(x) \sim x^{T - 1}$. Requiring that  $P(x)$ is continuous at $x^{\rm{typ}}$ shows then that
the left piece of the distribution is given by
          \begin{align}
            P_{\rm{left}} (x) \approx B_N x^{T - 1}, \qquad B_N \sim  N^{1/2 - T}\ .
            \label{pxleft}
          \end{align}
          
\begin{figure}
            \centering
            \includegraphics[width=0.48\textwidth]{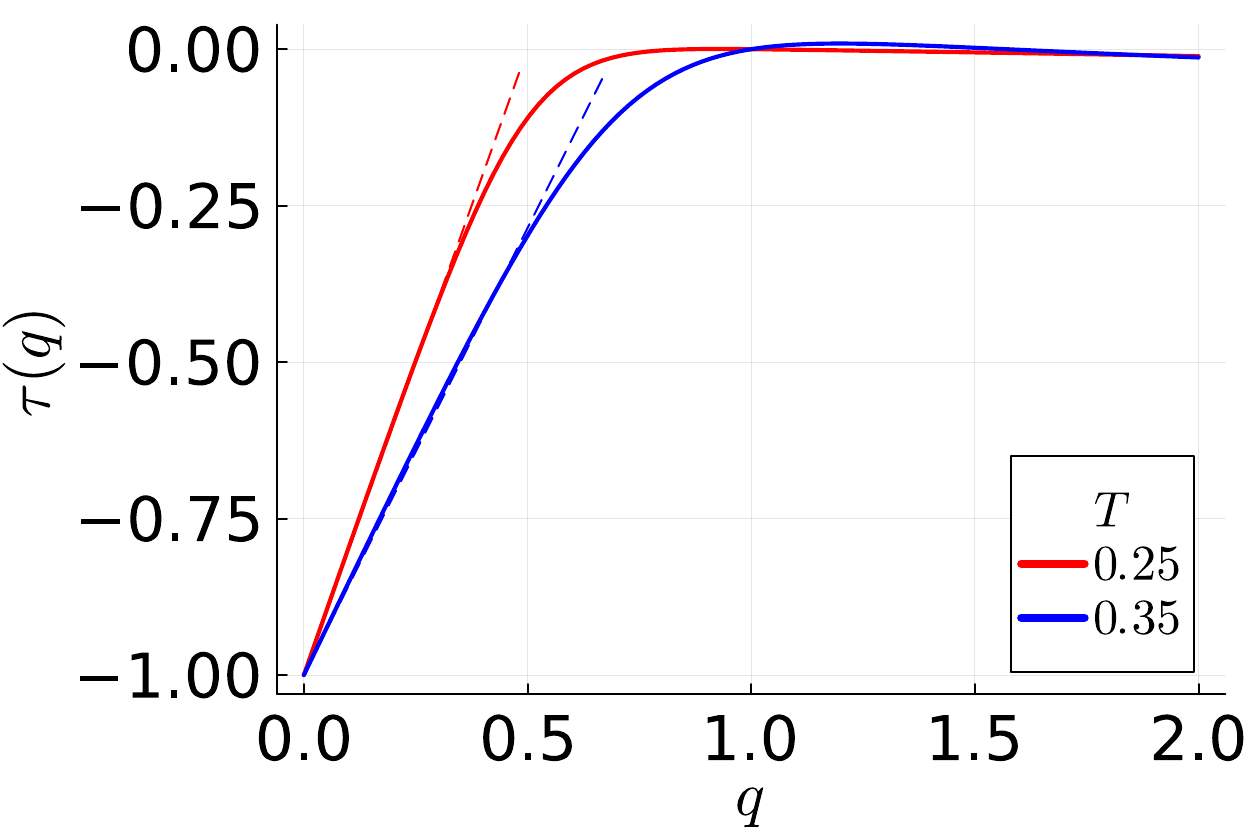} 
        \caption{Multifractal exponents $\tau(q)$ for $N = 2^{16}$ and two different temperatures. Dashed lines correspond to the analytical prediction given in Eq.~\eqref{moment_scaling}.}
        \label{mf_nb}
      \end{figure}

From $P(x)$ we can next obtain the set of multifractal exponents characterizing the moments of the distribution of wavefunction amplitudes~\cite{evers2008anderson, monthus2017statistical}. These moments and the associated exponents are defined by $I_q(N) = \sum_i |\psi_i|^{2q}  \propto N^{-\tau(q)}$. In terms of $x$ the moments are $I_q(N)=N\langle (x/N)^q\rangle$. The averages $\langle x^q\rangle$ can be found from~(\ref{px},\ref{typ_value}) for general $q>0$, in the same way as above for $q=0$ and $q=1$. They scale as $A_N(x^{\rm typ})^{q-2T}$ for $q<2T$ and as $A_N N^{q-2T}$ for larger $q$. Inserting into the definition of $I_q(N)$ lets one read off the multifractal exponents as  
           \begin{align}
             \tau(q) =
             \begin{cases}
               \frac{q}{2T} - 1\, , \qquad &q \leq 2T \\
               0  \qquad &q > 2T
             \end{cases}
             \label{moment_scaling}
           \end{align}
Fig.~\ref{mf_nb} shows the agreement of this expression with numerical data obtained by exact diagonalization (App.~\ref{ap1}), with the rounding around $q = 2T$ resulting from finite-size effects. This behavior reflects the localization of the wavefunctions, with eigenstates concentrated on a vanishing fraction of sites—an aspect we will characterize in the next section. A linear scaling of $\tau(q)$ followed by a plateau at $\tau(q) = 0$ has been observed in a wide range of systems exhibiting localization with multifractal properties~\cite{chou2014chalker, kravtsov2015random, tapias2023multifractality, da2025spectral}.

We note finally that an alternative way of capturing the multifractal properties of the wavefunctions is via the singularity spectrum (or spectrum of fractal dimensions) $f(\alpha)$. This is obtained by associating with each $x_i$ a fractal dimension via $x_i=N^{1-\alpha_i}$. The probability density of $\alpha$ is then obtained by variable transform from $P(x)$, and $f(\alpha)$ is defined via $P(\alpha)\sim N^{f(\alpha)-1}$.
Summarizing, $f(\alpha)$ can be obtained from
the formula~\cite{de2014anderson, kravtsov2015random, monthus2017statistical}
 \begin{align}
   f(\alpha) = \lim_{N\to  \infty} \frac{\ln (x N P(x))}{\ln N}
 \end{align}
where on the r.h.s.\ $x = N^{1 - \alpha}$.
From equations~\eqref{px} and~\eqref{pxleft} one then finds
           \begin{align}
             f(\alpha) =
             \begin{cases}
               2T \alpha \, , \qquad &0 \leq \alpha \leq 1/(2T)  \\
               3/2 - T\alpha  \, , \quad &1/(2T) \leq \alpha \leq 3/(2T)
             \end{cases}
             \label{mfspectrum}
           \end{align}
           i.e.\ the singularity spectrum has a triangular shape with slope $2T$ on the left and $-T$ on the right. A comparison with numerical results from exact diagonalization is shown in Fig.~\ref{fig_singularity} in App.~\ref{ap1}. From $f(\alpha)$ one can go back to the multifractal exponents via the Legendre transform~\cite{halsey1986fractal, monthus2017statistical} 
           \begin{align}
             -\tau(q) = \max_\alpha\,[f(\alpha) - q\alpha]
             \label{legendre}
           \end{align}
and straightforward algebra shows that this indeed retrieves~\eqref{moment_scaling}. 
Note that as seen in other models~\cite{da2025spectral}, the right wing of $f(\alpha)$, corresponding to the left tail~\eqref{pxleft} of $P(x)$, does not show up in $\tau(q)$ for $q>0$ and would only play a role if one were to consider $q<0$.

 
      \section{Effective Model}

So far we have established numerically that the wavefunctions (essentially relaxation modes) of the sparse BM model, at its first nonzero spectral divergence ($\lambda=-1/c$), have interesting multifractal properties that differ from the standard Anderson localization scenario. 
But we do not yet have an explanation for the form of the divergent spectral density (eq.~\eqref{specd}) for $T<1/2$, nor for the power--law tails of the distribution of squared wavefunction amplitudes (eqns.~\eqref{px} and~\eqref{pxleft}). Our aim in the rest of this paper is to construct a simplified effective model that allows us to rationalize these results and to understand the nature of localization in the wavefunctions concerned.

We construct this effective model around the insights set out in Sec.~\ref{divergences_recap}, primarily that the spectral divergences at for $\lambda=0$, $-1/c$, \ldots are remnants of the delta-functions that make up the spectral density for $T\to 0$. In particular, the spectral divergence at $\lambda=0$ corresponds physically to the slow, thermally activated escape out of low-lying local minima in the network of traps. Moving on to the first nonzero $\lambda=-1/c$, we saw that this eigenvalue for $T\to 0$ corresponds to transitions from saddles of index 1 into their one neighboring trap that is at larger depth. At nonzero $T$, additional transitions to shallow traps do of course become possible and will change $\lambda$, but these effects are expected to remain weak as long the saddle itself is low in the energy landscape, i.e.\ has large $E$. We therefore focus in the effective model on such deep saddles, which apart from their lower-lying neighbour are surrounded by shallow traps. Formally we define some large lower threshold $E_c$ on trap depth. We then select from the full network of traps the saddles ($s$) that lie at depth $E_s>E_c$ and that have exactly one deeper neighbour, with both being otherwise connected to shallow traps ($E<E_c$). The deeper neighbour is then necessarily a local minimum ($m$) at depth $E_m>E_s$, and together with the saddle it forms what we will call a ``saddle--minimum pair''. We write $P$ for the total number of such pairs and $(s_p,m_p)$ for the nodes in the $p$-th pair, with $p=1,\ldots,P$. 
%
         \begin{figure}
            \centering
            \includegraphics[width=0.65\textwidth]{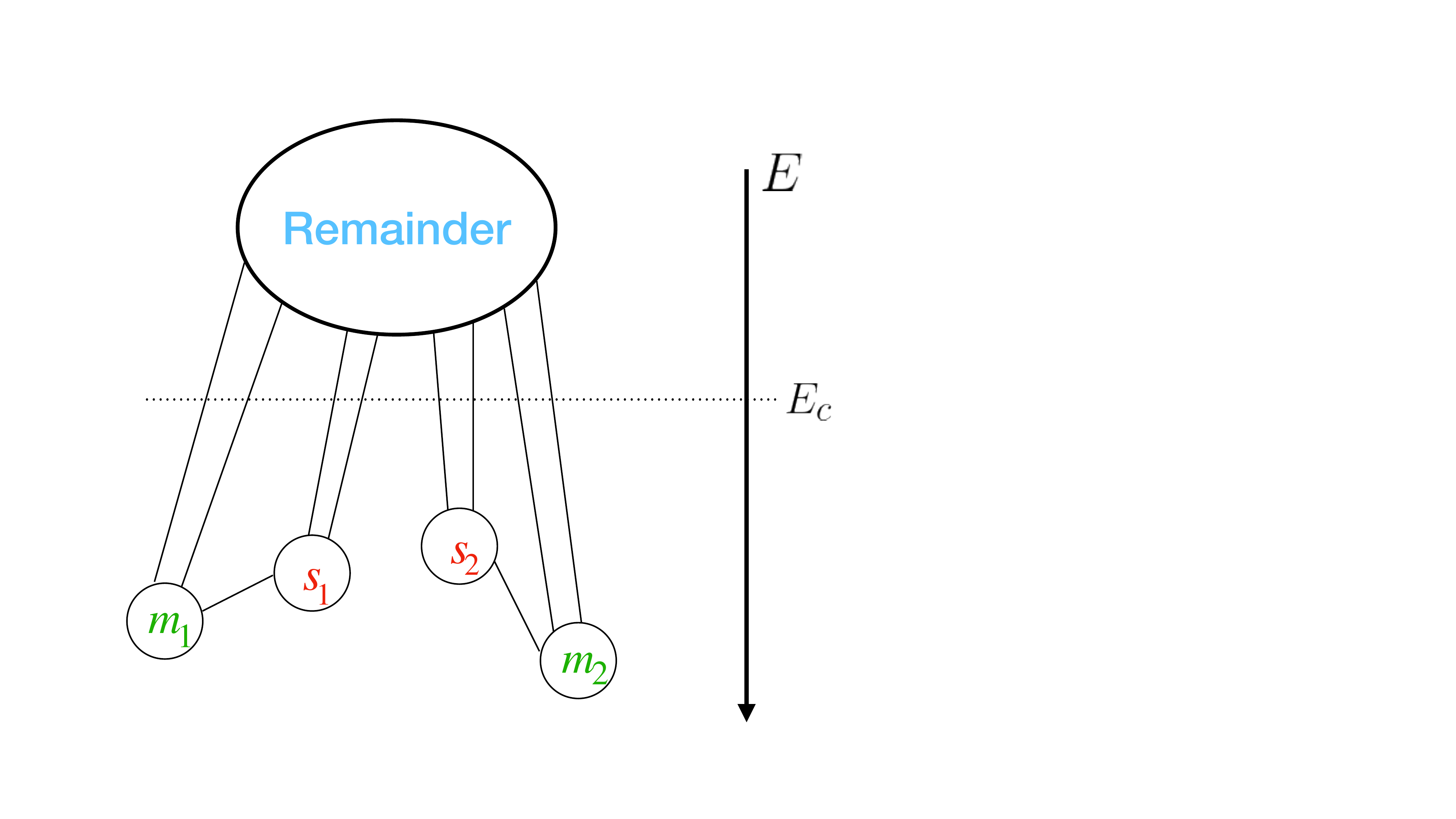} 
            \caption{Schematic of the effective model. We identify low-lying (at depth $>E_c$) node pairs consisting of an index 1 saddle and a local minimum, labelled ($s_1,m_1$) and ($s_2,m_2$) in the figure. The remainder of the network nodes is grouped into a single effective node, to which each saddle--minimum pair is connected by $c-1$ links from both the saddle and the minimum; in the figure, $c=3$.
}
        \label{eff_model}
      \end{figure}

To finish the effective model we now make a timescale separation approximation: we assume that $E_c$ is large enough so that thermally activated escapes from any saddle--minimum pair are slow compared to the dynamics among the $\nr=N-2P$ nodes in the remainder of the network. We therefore lump these remaining nodes into a single effective remainder node ($R$), to which each saddle--minimum pair is connected. 
Fig.~\ref{eff_model} illustrates the structure of the resulting effective network.


The corresponding effective master operator $\tilde{\Mm}$ contains transition rates between the saddle and minimum of each pair, of the unapproximated Glauber form~\eqref{mast}:
      \begin{eqnarray}
        \Wto{s}{m} &=& \frac{1}{c} \frac{1}{1 + \exp(-\beta(E_m - E_s))}
        \label{W_s_to_m} \\
        \Wto{m}{s} &=& \frac{1}{c} \frac{1}{1 + \exp(-\beta(E_s - E_m))}
        \label{W_m_to_s}
      \end{eqnarray}
with the same index convention as previously for $M_{ji}$, i.e.\ $\Wto{s}{m}$ is the rate for transitions from saddle to minimum. To the remainder node we associated an effective energy depth of $E_R=0$, consistent with the assumption of fast dynamics within the remainder, so that the transition rates from any saddle or minimum to the remainder are
\begin{eqnarray}
\Wto{s}{R} &=& \frac{c-1}{c} \frac{1}{1 + \exp(\beta E_s )}
        \label{trans_rem2}
\\
\Wto{m}{R} &=& \frac{c-1}{c} \frac{1}{1 + \exp(\beta E_m)}
\label{trans_rem2b}
\end{eqnarray}
The prefactor $c-1$ accounts for the fact that in the original network, each saddle and each minimum has $c-1$ connections to the remainder of the network. Finally for transitions from the remainder to a node within a saddle--minimum pair we assign the rates
      \begin{align}
        \Wto{R}{s} &= \frac{c-1}{c \nr} \frac{1}{1 + \exp(-\beta E_s))}
        \label{trans_rem1} \\
        \Wto{R}{m} &= \frac{c-1}{c \nr} \frac{1}{1 + \exp(-\beta E_m))}
        \label{trans_rem1b}
%
      \end{align}
The additional prefactor $1/\nr$ comes from the assumption that dynamics within the remainder of the network is fast and takes place at energy depth $E_R=0$, so that the probability of occupying a given (original) remainder node is uniform at $1/\nr$. The probability of occupying one of the $c-1$ (original) remainder nodes connected to a specific saddle or minimum is then $(c-1)/\nr$, giving~(\ref{trans_rem1},\ref{trans_rem1b}). 
It is easy to verify that the effective model as defined above still obeys detailed balance with respect to the Boltzmann distribution: the Boltzmann probabilities are $({\rm e}^{-\beta E_s}/Z,{\rm e}^{-\beta E_m}/Z)$ for each saddle-minimum pair, and $\nr/Z$ for the effective remainder node.

The effective model is still disordered, via the random trap depths of the saddles and minima. The number $P$ of saddle--minima pairs is in principle also a random quantity, but for large $N$ it will be self-averaging. To calculate it, note that 
the probability of a trap having $E>E_c$ is $\int_{E_c}^\infty dE\,\rho_E(E)=\exp(-E_c)$. The probability of a randomly chosen pair of neighbouring nodes to form a saddle--minimum pair is then
      \begin{align}
        \pi = [\exp(-E_c)]^2[1 - \exp(-E_c)]^{2(c-1)}
        \label{pi}
      \end{align}
where the second factor accounts for the requirement that all surrounding nodes, i.e.\ all $c-1$ other neighbours of both the saddle and the minimum, should be shallow traps. Given that the total number of node pairs is $cN/2$, the number of saddle--minimum pairs in a large network is then $P=(c/2)N\pi$ and the remainder of the network will contain $\nr=N-2P=N(1-c\pi)$ nodes.

To find the joint probability distribution of the trap depth $E_s$ and $E_m$ of, respectively, the saddle and the minimum in a pair, we note that both are drawn from $\rho_E(E)$ but conditioned on $E>E_c$. In addition the depths are ordered, $E_m>E_s$, so that 
      \begin{align}
    P(E_s, E_m) =  2\, {\rm{e}}^{-(E_m - E_c)  }  {\rm{e}}^{-(E_s - E_c)  }  \,\Theta(E_m  - E_s)\,\Theta(E_s  - E_c)
\label{joint_saddle_minimum}
      \end{align}
with $\Theta( \cdot)$ the Heaviside step function. For later it will be more convenient to rewrite this in terms of the saddle depth and the energy gap $\Delta E=E_m -E_s$ as
      \begin{align}
    P(E_s , \Delta E) =  2\, {\rm{e}}^{- \Delta E} {\rm{e}}^{-2 (E_s - E_c)}\,\Theta(\Delta E)\,\Theta(E_s  - E_c)
    \label{joint_saddle_delta}
  \end{align}

  To conclude our presentation of the effective model we comment briefly on the role of deep local minima, i.e.\ nodes at depth $E>E_c$ that are surrounded by shallow nodes. There are $N\exp(-E_c)[1-\exp(-E_c)]^{c-1}$ of these in the network, which for large $E_c$ is significantly larger than the number of nodes in the $P$ saddle--minimum pairs, due to the $\exp(-2E_c)$ factor in~\eqref{pi}. One can indeed extend the effective model to include these deep local minima, but -- consistent with the arguments above for the spectral divergence at $\lambda=0$ -- they only contribute to the spectrum for small $\lambda$ and have a negligible effect on the spectrum around $\lambda=-1/c$. We therefore proceed without including the deep local minima explicitly (see Ref.~\cite{takaki2023theory} for an effective model with (only) the minima and one remainder node, constructed there in the context of aging in protein condensates).

  \subsection{Spectral density}

We can now proceed to finding the spectral density of the effective model, focussing on the region around $\lambda=-1/c$ where we expect a divergence in $\rho(\lambda)$.

The effective master operator $\tilde{\Mm}$ is a $(2P+1)\times(2P+1)$ matrix. We arrange its rows and columns by listing the saddle minimum pairs in order ($s_1,m_1,s_2,m_2,\ldots,s_P,m_P$), followed by the remainder node $R$ at position $2P+1$. The matrix therefore has the structure of a block diagonal matrix formed by $P$  blocks $\Btm_p$ (of size $2\times 2$, one for each pair $p$), augmented by a final row and column containing the transition rates to and from the remainder node (see equation~\eqref{modmaster} below). The structure of each block is given by 
      \begin{align}
        \Btm_p = 
        \begin{pmatrix}
          - \Gamma_{s_p} & \Wto{m_p}{s_p} \\
          \Wto{s_p}{m_p}  & - \Gamma_{m_p}
        \end{pmatrix}
      \end{align}
where on the diagonal we have the escape rates from the saddle or minimum, respectively, which are
      \begin{align}
        \label{esc_saddle}
        \Gamma_s &= \Wto{s}{m} + \Wto{s}{R} \\
        \Gamma_m &= \Wto{m}{s} + \Wto{m}{R}
                   \label{esc_minimum}
      \end{align}
Here and below we drop the subscript $p$ for a generic saddle--minimum pair when convenient, in order to lighten the notation. The whole matrix can now be expressed as
      \begin{align}
        \tilde{\Mm} = 
              \begin{pNiceMatrix}[margin]
\Block{2-2}{\Btm_1} & & 0 & 0 & 0 & 0 &  \Wto{R}{s_1} \\
                 & & 0 & 0 & 0 & 0 & \Wto{R}{m_1} \\
0 & 0 & \Block{2-2}{\ddots} & & 0 & 0 &  \Block{2-1}{\vdots}  \\
0 & 0 &                  & & 0 & 0  & \\
0 & 0 & 0 & 0 & \Block{2-2}{\Btm_P}  & & \Wto{R}{s_P} \\
0 & 0 & 0 & 0  & & & \Wto{R}{m_P} \\
\Wto{s_1}{R} & \Wto{m_1}{R} & \Block{1-2}{\cdots} & & \Wto{s_P}{R} & \Wto{m_P}{R} &-\Gamma_{R}
\end{pNiceMatrix}
        \label{modmaster}
      \end{align} 
As for the full model (see Sec.~\ref{sec:bm}) we can symmetrize the matrix $\tilde{\Mm}$ using detailed balance w.r.t.\ the Boltzmann weights. From~\eqref{M_symm} one sees that $\tilde{M}^s_{ij}\tilde{M}^s_{ji}=
P_{{\rm eq},i}^{-1/2} \tilde{M}_{ij} P_{{\rm eq},j}^{1/2} P_{{\rm eq},j}^{-1/2}  \tilde{M}_{ji}P_{{\rm eq},i}^{1/2} = \tilde{M}_{ij}\tilde{M}_{ji}$, giving a simple way to calculate the symmetrized matrix as
      \begin{align}
        \tilde{M}^s_{ij} = \sqrt{\tilde{M}_{ij} \tilde{M}_{ji}}
      \end{align}

To find the spectrum, the key property of the effective master operator~\eqref{modmaster} and its symmetrized version $\tilde{\Mm}^s$ is that it can be written as a low--rank (specifically, rank 2) perturbation of a block diagonal matrix. The latter, which we will call $\Dm$, contains the $P$ symmetrized $2\times 2$ blocks and,  at position $(2P+1, 2P+1)$, the negative escape rate from the remainder, $-\Gamma_R$. Each symmetrized block, denoted below by $\Bm_p$, has the form
        \begin{align}
        \Bm =
        \begin{pmatrix}
          - \Gamma_{s} & \sqrt{\Wto{s}{m} \Wto{m}{s}} \\
          \sqrt{\Wto{s}{m} \Wto{m}{s}}  & - \Gamma_{m}
        \end{pmatrix}
        \label{block_bs}
      \end{align}
where we have again dropped the subscript $p$.
On the other hand the rank--2 perturbation matrix, denoted by $\Qm$, can be written in terms of the following $(2P+1)$--dimensional column vectors:
      \begin{align}
        \bm{w} &= (\sqrt{\Wto{R}{s_1}  \Wto{s_1}{R} }, \ldots, \sqrt{\Wto{R}{m_P}  \Wto{m_P}{R} }, 0)\T \label{w1} \\
        \bm{e} &= (0, 0, \ldots, 0, 1)\T
                 \label{e1}
      \end{align}
The whole symmetrized master operator is then given by
      \begin{align}
        \tilde{\Mm}^s  = \Dm +      \underbrace{\bm{w}    \bm{e}\T +       \bm{e}  \bm{w}\T}_{ \Qm }
        \label{wholem}
      \end{align}
      
We next use the above decomposition to show that the spectral density of $\tilde{\Mm}^s$ is essentially identical to that of $\Dm$ for large $N$. 
We start by writing the characteristic equation  as
      \begin{align}
        \det (\lambda  {\id} -            \tilde{\Mm}^s ) =   \det (\lambda  {\id} -            {\Dm}  ) \det ( \id -         (\lambda  {\id} -            {\Dm} )^{-1}  {\Qm}  ) \overset{!}{=} 0
      \end{align}
The first factor is nonzero unless an eigenvalue $\lambda$ of $\tilde{\Mm}^s$ is also an eigenvalue of $\Dm$, a case we will be able to exclude below. The eigenvalue condition is then equivalent to 1 being an eigenvalue of the matrix $  (\lambda  {\id} -            {\Dm} )^{-1}  {\Qm} $, which means there must be a (nonzero) vector $\uv$ obeying
      \begin{align}
        (\lambda  {\id} -            {\Dm} )^{-1}  {\Qm}  \uv = \uv\ .
        \label{u_cond}
      \end{align}
Multiplying both sides by $ {\Qm} $ and using the definition of this matrix as given~\eqref{wholem} shows that 
\begin{equation}
{\Qm}  \uv = \gam_w \bm{w}  + \gam_e \bm{e}
\label{psi_representation}
\end{equation}
with $\gam_w$ and $\gam_e$  coefficients to be found. Defining
\begin{align}
        {\Xm} =   (\lambda  {\id} -            {\Dm} )^{-1}
        \label{Xdef}
      \end{align}
we can thus reduce the eigenvalue condition $\Qm \Xm \Qm \uv = \Qm \uv$
to the $2 \times 2$ system 
      \begin{align}
        \begin{pmatrix}
\bm{e}\T {\Xm} \bm{w} & \bm{e}\T {\Xm} \bm{e}  \\
\bm{w}\T {\Xm} \bm{w} & \bm{w}\T {\Xm} \bm{e} 
        \end{pmatrix}
        \begin{pmatrix}
          \gam_w \\
          \gam_e
        \end{pmatrix}
        =
            \begin{pmatrix}
          \gam_w \\
          \gam_e
            \end{pmatrix}
        \label{evalue_sys}
      \end{align}
Because of the block diagonal structure of $\Xm$, the vector $\Xm\bm{e}$ only has one nonzero entry, at the last position $2P+1$, and is therefore orthogonal to $\bm{w}$. The diagonal entries in the matrix on the left of~\eqref{evalue_sys} then vanish and one is left with
      \begin{align}
        \label{cond1}
      \bm{e}\T {\Xm} \bm{e}\, \gam_e = \gam_w\\
      \bm{w}\T {\Xm} \bm{w}\, \gam_w = \gam_e
        \label{cond2}
      \end{align}
Hence there exists a solution of the system~\eqref{evalue_sys} if and only if
      \begin{align}
        (  \bm{w}\T {\Xm} \bm{w})  (\bm{e}\T {\Xm} \bm{e})  = 1
        \label{condw}
      \end{align}
Writing this out and using the definition~\eqref{e1} of $\bm{e}$ gives
      \begin{align}
        \frac{1}{\lambda + \Gamma_R}            \sum_{i,j= 1}^{2P}  w_i X_{ij} w_j = 1
        \label{eigvalue}
      \end{align}
Now if we denote the eigenvalues of $\Dm$ by $\lambda_i^{\Dm}$, then the eigenvalues of $\Xm$ are $(\lambda-\lambda_i^{\Dm})^{-1}$. Rewriting the sum in~\eqref{eigvalue} in the eigenbasis of $\Xm$, with $\tilde{w}_i$ the correspondingly transformed elements of $\bm{w}$, we thus have
      \begin{align}
         \sum_{i=1}^{2P}  \frac{\tilde{w}_i^2}{\lambda - \lambda_i^{\Dm}} = \lambda + \Gamma_R           
        \label{fin_evalue}
      \end{align}
where the eigenvalues $\lambda_1^{\Dm}, \ldots,  \lambda_{2P}^{\Dm}$ are the ones associated with the $P$ saddle--minima blocks of $\Dm$. (The remaining eigenvalue is $\lambda_{2P+1}^{\Dm} = -\Gamma_R$.) 
Given that the l.h.s.\ of~\eqref{fin_evalue} is a succession of poles with positive weights, one reads off that there must be exactly one eigenvalue of $\tilde{\Mm}^s$ between any pair of adjacent $\lambda_i^{\Dm}$. 
This interlacing of eigenvalues (see e.g.~\cite{horn2012matrix}) 
implies -- as claimed above -- that the matrices $ \tilde{\Mm}^s$ and $\Dm$ have the same spectral density in the limit $N\to \infty$, where the distance between adjacent eigenvalues of $\Dm$ shrinks to zero.

The remaining task is to calculate the first $2P$ eigenvalues of $\Dm$. Because of the block diagonal structure of the matrix this eigenvalue calculation reduces to separate eigenvalue problems for each of its constituent blocks $\Bm$, the latter having been defined in equation~\eqref{block_bs}. The corresponding characteristic equation reads
      \begin{align}
        (\lambda + \Gamma_s)(\lambda + \Gamma_m) - \Wto{s}{m} \Wto{m}{s} = 0
      \end{align}
and gives the eigenvalues
      \begin{align}
        \lambda = - \frac{(\Gamma_m + \Gamma_s) \pm \sqrt{(\Gamma_m + \Gamma_s)^2  - 4( \Gamma_m \Gamma_s - \Wto{s}{m} \Wto{m}{s})} }{2}
        \label{full_lambda}
      \end{align}
From here we can in principle obtain the spectrum of $\Dm$ for $N\to\infty$ by drawing saddle--minimum pairs $(E_s,E_m)$ according to~\eqref{joint_saddle_minimum}, which determine the rates~(\ref{W_s_to_m}--\ref{trans_rem2b}) and so the eigenvalues $\lambda$. This does not yield a closed form for the spectrum, however, so we simplify first by exploiting the fact that both the saddle and the minimum are deep in the energy landscape, $E_s$, $E_m>E_c$. The rates~(\ref{trans_rem2},\ref{trans_rem2b}) for transitions up to the remainder node are then exponentially small, being bounded by $\exp(-\beta E_c)$. Expanding to first order in these Arrhenius factors gives after a short calculation
\begin{eqnarray}
\lambda^{(1)} &=&
{}-(\Wto{s}{m}+\Wto{m}{s})-\frac{\Wto{s}{m}\Wto{s}{R}+\Wto{m}{s}\Wto{m}{R}}{\Wto{s}{m}+\Wto{m}{s}}
\\
\lambda^{(2)} &=& {}-\frac{\Wto{s}{m}\Wto{m}{R}+\Wto{m}{s}\Wto{s}{R}}{\Wto{s}{m}+\Wto{m}{s}}
\end{eqnarray}
Now the sum of the transition rates within the pair is $\Wto{s}{m}+\Wto{m}{s}= 1/c$: it is therefore (only) the first eigenvalue that will contribute to the spectrum around $\lambda=-1/c$ while the second eigenvalue is close to zero. Explicitly we have, dropping the (1) superscript,
%
      \begin{eqnarray}
        \lambda+1/c    &=& - c (
        \Wto{s}{m}\Wto{s}{R} +
        \Wto{m}{s}\Wto{m}{R} )
                        \label{lambda1}\\
                        &=& -\frac{c-1}{c}\,
\frac{1+{\rm e}^{-2\beta \Delta E}}{1 + {\rm e}^{-\beta\Delta E}} {\rm{e}}^{-\beta E_s}
            \label{zeta}
\end{eqnarray}
where in the second line we have explicitly inserted the Glauber rates~(\ref{W_s_to_m},\ref{W_m_to_s}) and the escape rates~(\ref{trans_rem2},\ref{trans_rem2b}), expanded to first order in Arrhenius factors. The $\Delta E$-dependent fraction is $O(1)$ (and in fact bounded between $2(\sqrt{2}-1)\approx 0.828$ and $1$). Calling $z=\lambda+1/c$, the distribution of $z$ for small $z$ is thus governed by the 
exponential distribution~\eqref{joint_saddle_delta} of saddle depths $E_s$, giving 
      \begin{align}
        \rho(z) \sim |z|^{2T-1}
        \label{rho_zeta}
      \end{align}
Our effective model thus allows us understand the observed spectral divergence~\eqref{specd} of the full model including the exponent. The physical intuition can be summarized as follows: eigenvalues $\lambda$ near $-1/c$ arise from low-lying saddle--minimum pairs, with the difference $z=\lambda-1/c$ determined by an Arrhenius factor. This gives a spectral divergence similar to the one at $\lambda=0$ but with the crucial difference that saddle--minimum pairs, requiring the presence of two adjacent low-lying traps, have a lower density of states $P(E_s)\propto [\rho_E(E_s)]^2$.

\subsection{Eigenvector statistics}

We next examine the statistics of the eigenvector entries for the effective model, focussing on eigenvalues with eigenvalues at the spectral divergence $\lambda=-1/c$.

The eigenvector condition~\eqref{u_cond} can be written as      
      \begin{align}
        \uv =  (\lambda  {\id} -            {\Dm} )^{-1}  {\Qm}  \uv = {\Xm} (\gam_w \bm{w}  + \gam_e \bm{e})
        \label{u2}
      \end{align}
using the representation~\eqref{psi_representation} and the definition~\eqref{Xdef}. The coefficients $\{\gam_w, \gam_e\}$ satisfy the system~(\ref{cond1},\ref{cond2}) and are thus proportional to each other;
eliminating $\gam_e$ using~\eqref{cond1} yields explicitly
      \begin{align}
        \uv = 
\gam_w  \left(  {\Xm}  \bm{w} +  \frac{{\Xm}  \bm{e}}{ \bm{e}\T {\Xm} \bm{e} } \right)
= \gam_w  \left(  {\Xm}  \bm{w} +  \bm{e} \right)
        \label{evector}
      \end{align}
where the last equality follows from the definition~\eqref{e1} of the vector $\bm{e}$.  
The second term in brackets in~\eqref{evector} contributes only to the last entry (at position $2P+1$) of $\uv$.
The first $2P$ entries of $\uv$ 
can be found separately for each of the saddle--minimum pairs $p=1,\ldots, P$, due to the block structure of $\Xm$. For a generic pair the components of $\uv$ are, from~\eqref{evector},
  \begin{equation}
\gam_w^{-1}    \left(\!\!
    \begin{array}{c}
    \u_{s} \\ \u_{m} 
    \end{array}
\!\!\right)   =
    (\lambda  {\id} -            \Bm )^{-1}  
        \left(\!\!
    \begin{array}{c}
    w_{s} \\ w_{m} 
    \end{array}
\!\!\right)   
\label{psi_pair}
  \end{equation}
where $(w_s,w_m)$ are the corresponding components of $\bm{w}$ and $\Bm$ is as defined in~\eqref{block_bs}. Evaluating the inverse explicitly and inserting $\lambda = -1/c = -(\Wto{s}{m}+\Wto{m}{s})$ gives
\begin{eqnarray}
(\lambda\id-\Bm)^{-1}=
\frac{1}{D}\left(
\begin{array}{cc}
\Wto{s}{m}-\Wto{m}{R} &
-\sqrt{\Wto{m}{s}\Wto{s}{m}}
\\
-\sqrt{\Wto{m}{s}\Wto{s}{m}} & 
\Wto{m}{s}-\Wto{s}{R} 
\end{array}\right)
\label{inverse_explicit}
\end{eqnarray}
with the determinant
\begin{equation}
D = \Wto{s}{m}\Wto{s}{R}
+\Wto{m}{s}\Wto{m}{R}-\Wto{m}{R}\Wto{s}{R}\ .
\label{determinant}
\end{equation}
As before we now retain only the leading in the small Arrhenius factor $\Wto{s}{R}$ and $\Wto{m}{R}$, i.e.\ we neglect the negative terms on the diagonal in~\eqref{inverse_explicit} and the last term in~\eqref{determinant}. Within this approximation one has
\begin{eqnarray}
(\lambda\id-\Bm)^{-1}&= &
\frac{1}{D}\left(
\begin{array}{cc}
\Wto{s}{m} &
-\sqrt{\Wto{m}{s}\Wto{s}{m}}
\\
-\sqrt{\Wto{m}{s}\Wto{s}{m}} & 
\Wto{m}{s}
\end{array}\right)\\
D &= & \Wto{s}{m}\Wto{s}{R}
+\Wto{m}{s}\Wto{m}{R}
\end{eqnarray}
and so from~\eqref{psi_pair} 
\begin{eqnarray}
\gam_w^{-1}\left(\!\!\begin{array}c\u_s\\\u_m
\end{array}\!\!\right)
= \frac{1}{D}\left(\begin{array}{c}
\Wto{s}{m}w_s
-\sqrt{\Wto{m}{s}\Wto{s}{m}}w_m\\
-\sqrt{\Wto{m}{s}\Wto{s}{m}} w_s + 
\Wto{m}{s}w_m\end{array}
\right)
\label{psi_explicit}
\end{eqnarray}
Taking the ratio of the two vector components, one reads off
\begin{equation}
\u_m = -{}\u_s \sqrt{\Wto{m}{s}/\Wto{s}{m}} = {}- {\rm{e}}^{- \beta  \Delta E/2 } \u_s   
\label{minimum-saddle}
\end{equation}
using the explicit expression for the rates or detailed balance
$\Wto{m}{s}/\Wto{s}{m} = 
{\rm{e}}^{\beta (E_s-E_m)} = 
{\rm{e}}^{- \beta  \Delta E} 
$. This result makes sense as it is exactly what one would find for the relaxation mode in a saddle--minimum pair that is disconnected from the remainder of the network, i.e.\ in the limit where the Arrhenius factors are zero.

It now remains to find $\u_s$. This can be simplified using that, to leading order in the Arrhenius factors,
\begin{equation}
\Wto{s}{R}=\frac{c-1}{c}{\rm{e}}^{-\beta E_s}, \qquad 
w_s = \frac{c-1}{c \nr^{1/2}}{\rm{e}}^{-\beta E_s/2}
\end{equation}
and similarly for transitions out of the minimum (with $E_s$ replaced by $E_m$). Inserting these expressions into~\eqref{psi_explicit} yields
\begin{eqnarray}
\gam_w^{-1} \psi_s &=& 
\nr^{-1/2} \frac{\Wto{s}{m}{\rm{e}}^{-\beta E_s/2}-\Wto{s}{m}{\rm{e}}^{-\beta \Delta E/2}{\rm{e}}^{-\beta E_m/2}}{\Wto{s}{m}  {\rm{e}}^{-\beta E_s}
+\Wto{s}{m} {\rm{e}}^{-\beta \Delta E}{\rm{e}}^{-\beta E_m}
}
\\
&=&\nr^{-1/2} {\rm{e}}^{\beta E_s/2}\frac{1-{\rm{e}}^{-\beta \Delta E}}{1+{\rm{e}}^{-2\beta \Delta E}}
\label{saddle}
\end{eqnarray}

We can now finally use the joint distribution~\eqref{joint_saddle_delta} of $E_s$ and $\Delta E$ to obtain the distribution of the squared eigenvector entries ($x_s = N\u_s^2$ and $x_m=N\u_m^2$). 
Given that the fraction in~\eqref{saddle} is $O(1)$ for typical $\Delta E$, the eigenvector weights $x_s=N\u_s^2$ scale with ${\rm e}^{\beta E_s}$. Given that $P(E_s)\propto {\rm e}^{-2E_s}$, the result is a power--law tail in the distribution of $x$ resulting from deep saddles: 
  \begin{align}
    P_{s}(x) \sim 
    x^{-(2T + 1)}
    \label{psaddle}
  \end{align}
This is exactly the large-$x$ power--law tail that we had previously identified numerically~\eqref{px}. 
Because of the relation~\eqref{minimum-saddle}, the minima contribute with the same tail for large $x$. But they turn out to dominate the distribution for small $x$ because the wavevector entries on deep minima (large $\Delta E$) can become exponentially small from~\eqref{minimum-saddle}, $x_m \propto \u_m^2 \propto {\rm e}^{-\beta \Delta E}$. From $P(\Delta E)\propto {\rm e}^{-\Delta E}$ one then finds
  \begin{align}
    P_m(x) 
    \sim x^{T-1}
    \label{pmx}
  \end{align}
which accords perfectly with the numerically determined left wing of the squared wavefunction amplitude distribution (eq.~\eqref{pxleft}).

We note briefly that from~\eqref{saddle}, both saddle and minimum wavefunction entries do also become small for nearly degenerate saddle--minimum pairs ($\Delta E \to 0$): in that limit both $\u_s^2$ and $\u_m^2$ are proportional to $(\Delta E)^2$. This gives a small-$x$ tail of $P(x)\sim x^{-1/2}$. We have ignored this above because it is subleading compared to~\eqref{pmx} in the temperature regime $T<1/2$ that is of interest here because a spectral divergence actually appears. We have also not commented explicitly on the normalization factor $\gamma_w$, for which also the remainder component $\u_{2P+1}$ of the wavefunction needs to be taken into account. We show in App.~\ref{sec:normalization}, however,
that this remainder makes a subleading contribution to the normalization for low $T$. It therefore does not affect the conclusion that wavefunctions are concentrated onto deep saddles and the associated minima.

  \section{Discussion}

  In this work, we have explored the spectral statistics of the sparse Barrat--M\'ezard trap model around its unusual spectral divergences, focussing on the first nonzero such divergence at
$\lambda = -1/c$ with $c$ the network connectivity. To rationalize the numerically observed power laws 
in the 
spectral density and the 
tails of the squared wavefunction amplitude distribution we have also constructed a simplified effective model that is, to a large extent, analytically solvable.

Regarding the statistics of the eigenvectors of the symmetrized master operator, i.e.\ the relaxation modes, we found that the distribution of the squared eigenvector entries has the same qualitative features as the distribution of the local density of states with, in particular, temperature--dependent power law tails towards small and large values. 
We remark that the qualitative agreement between the statistics of the squared wavevector entries and the LDoS does not hold for Anderson-localized states on the Bethe lattice. There, the LDoS exhibits a right-sided power-law tail with a universal exponent $-3/2$~\cite{mirlin1994distribution}, whereas the statistics of eigenfunction amplitudes follows a disorder-dependent power-law~\cite{de2014anderson}, coinciding with the LDoS only at the critical point.

We characterized the localization properties of the eigenvectors by both the set of multifractal exponents $\tau(q)$ and the singularity spectrum $f(\alpha)$. The fact that 
$\tau(q) = 0$ above a certain $q^* = 2T$, or equivalently that the support of $f(\alpha)$ extends down to $\alpha_{\min} = 0$ with $f(\alpha_{\min}) = 0$, indicates eigenvectors that are concentrated on a finite set of nodes in the thermodynamic limit. Such states have in other contexts been denoted as ``frozen'', ``quasi--localized'' or localized with ``strong multifractal'' properties~\cite{evers2008anderson, chou2014chalker, kravtsov2015random, monthus2016localization, garciamata2022critical}. 
But in the sparse Barrat--M\'ezard model, this localization follows a pattern that is distinct from Anderson localization; we present further evidence of this from correlation functions and by direct visualization in Appendixes~\ref{app:corr} and~\ref{app:net}.
In fact, the unusual localization properties 
are closely related to a divergence in the bulk of the spectral density, a relation also seen in the scenarios of Refs.
~\cite{chou2014chalker, tapias2023multifractality, da2025spectral}. 

In Ref.~\cite{tapias2023multifractality} we introduced the concept of ``statistical localization'' to emphasize the distinct feature of localization occurring in systems like the sparse Barrat--M\'ezard model: it does not take place spatially (in the network sense) around a single node or set of nodes, but rather on nodes that are spatially widely separated but have similar statistical properties. In Ref.~\cite{tapias2023multifractality} the relevant node property was the relative degree. Here, we see from the analysis of our effective model that the relevant property is the depth of saddle nodes within deep saddle--minimum pairs. Indeed, the right power--law tail of the distribution of squared eigenvector entries (eq.~\eqref{px}) of the full model can be explained from the distribution of squared eigenvector entries within individual saddle--minimum pairs (eq.~\eqref{psaddle}), thus  supporting the interpretation in terms of statistical localization and providing a simple intuitive picture of the localization mechanism at work. 

A similarly intuitive picture also emerges from the effective model for the divergence of the spectral density: our calculation ultimately reveals that it is the thermal activation from deep saddles within a pair is responsible for the power--law divergence (eqs.~\eqref{zeta} and~\eqref{rho_zeta}). This is directly analogous to what happens at $\lambda \to 0$, where the spectral divergence arises from activation out of deep local minima, a fact that 
can be seen explicitly in the closely related Bouchaud trap model~\cite{margiotta2018spectral, riccardothesis}. The different exponent of the spectral divergence at $\lambda=-1/c$ arises because deep saddles are significantly rarer than deep minima, requiring the presence of two neighbouring nodes (rather than a single one) lying deep in the energy landscape. This is the origin of the characteristic temperature $T=1/2$, i.e.\ half the glass transition temperature, below which the spectral divergence appears. It would be 
interesting to explore whether the same intuition can be confirmed also for the spectral divergences at higher multiples of $-1/c$. Given that saddles of index two and higher would then be involved, multiple local network topologies of deep traps would need to be considered, which in turn might also explain the $c$-dependent power laws seen numerically~\cite{tapias2020entropic}.

It would also be worth extending the effective model calculation to the vicinity of the spectral divergence at $\lambda = -1/c$. This approach may help elucidate the role of the statistical localization mechanism in the Anderson-like phenomenology observed in our previous work~\cite{tapias2022localization}. Preliminary calculations indicate that for small but finite values of $z = \lambda + 1/c \neq 0$, the distribution of squared wavefunction amplitudes exhibits a power-law tail with exponent $-3/2$. This distribution is characteristic of Anderson-localized states on a Bethe lattice at the critical point~\cite{de2014anderson}, which highlights the relevance of its appearance in our context. Notably, the same tail also characterizes the localized states in the mean--field Bouchaud trap model~\cite{margiotta2018spectral}, as well as the localized eigenvectors of highly heterogeneous networks with homogeneous couplings, as analyzed in Ref.~\cite{da2025spectral}. A more detailed analysis is left for future work.

Finally, we note that, despite differences in the underlying physical mechanisms, our results bear a conceptual similarity to the recent study of multifractality in the weighted Erdős–Rényi (WER) ensemble~\cite{cugliandolo2024multifractal}, where non--trivial power laws in the local density of states (LDoS) were also observed and attributed to structural heterogeneity. In particular, we hypothesize that the divergent power-law behavior observed in the spectral density at $E = 0$ (in addition to the $\delta$-peaks associated with leaf nodes) reflects the presence of statistically localized states.

\appendix

\section{Numerical details and singularity spectrum}
\label{ap1}

The multifractal exponents and the singularity spectrum are estimated using the method described in Appendices C.1 and C.2 of Ref.~\cite{da2025spectral}. As there, we considered network sizes $N \in \{2^{15}, 2^{16}\}$. For both of these we generated instances of the master operator on random regular graphs with connectivity $c = 3$, according to Eqs.~(\ref{mast},~\ref{M_diag}).

\begin{figure}
  \centering
  \includegraphics[width=0.48\textwidth]{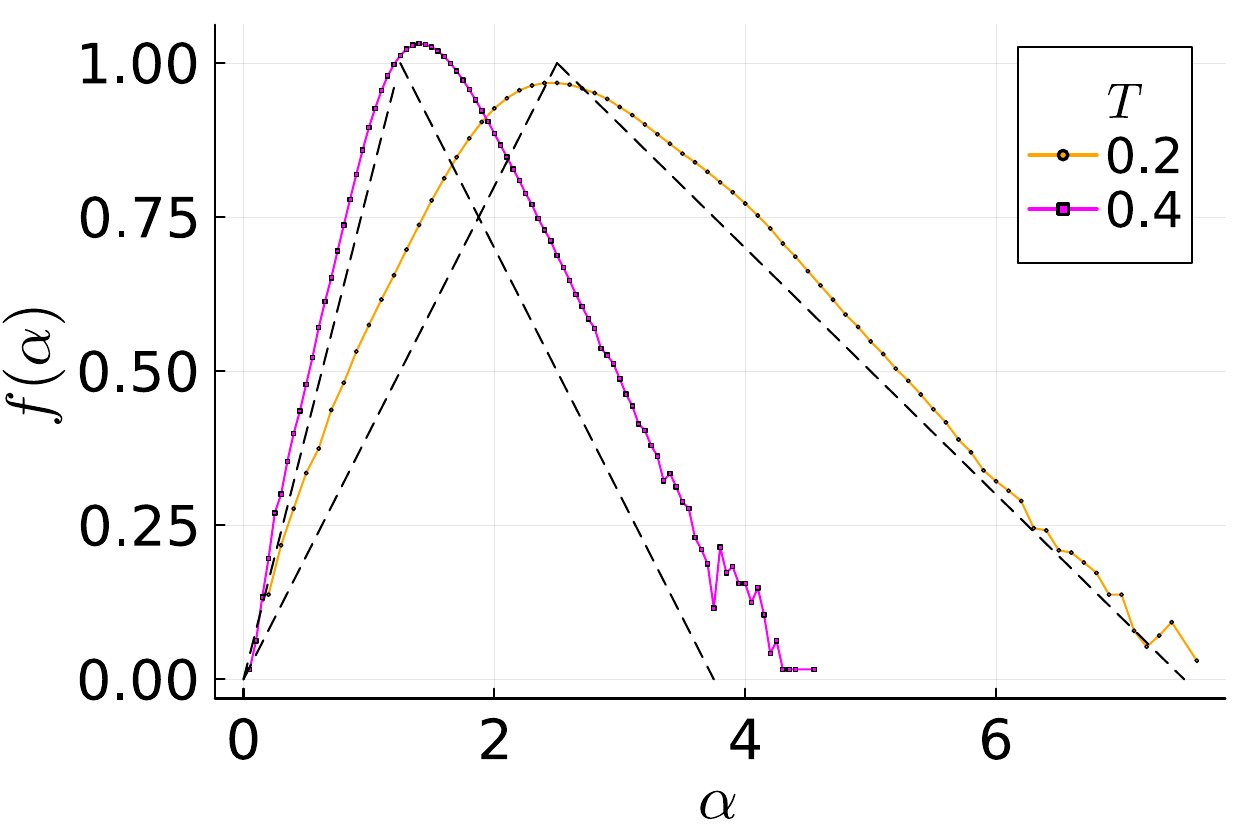}
  \caption{Singularity spectrum $f(\alpha)$ for $N = 2^{16}$ and two different temperatures. Dashed lines show the analytical prediction from Eq.~\eqref{mfspectrum}.}
  \label{fig_singularity}
\end{figure}

The resulting set of multifractal exponents is shown in Fig.~\ref{mf_nb}, while the singularity spectrum $f(\alpha)$ for $N = 2^{16}$ is shown in Fig.~\ref{fig_singularity}. As the figure indicates, the prediction from Eq.~\eqref{mfspectrum} is consistent with the numerical data, although rounding from finite-size effects remain visible.

\section{Normalization of eigenvectors}
\label{sec:normalization}

We briefly discuss the normalization condition for the eigenvectors of the effective master operator. The eigenvector component $\u_{2P+1}$, associated with the \emph{remainder} node  in the effective model, is given by $\u_{2P+1} = \gam_w$
from~\eqref{evector}.
Using~(\ref{minimum-saddle},\ref{saddle}) for the other eigenvector components gives the normalization condition
\begin{align}
\gam_w^{-2} &= 1+
\frac{1}{\nr}
\sum_{p=1}^{P}{\rm{e}}^{\beta E_{s_p}} \frac{(1 - {\rm{e}}^{-\beta \Delta E_p})^2(1 + {\rm{e}}^{-\beta \Delta E_p})}{(1 + {\rm{e}}^{-2\beta \Delta E_p})^2}\, .
\label{normalization}
\end{align}
As typical values of $\Delta E_p$ are of order unity, the same is true of the $\Delta E_p$-dependent factors in the sum. We therefore ignore these and estimate the sum by an integral over the distribution of $E_s$. The appropriate probability density follows from~\eqref{joint_saddle_delta} as
\begin{equation}
P(E_s)=2\, {\rm e}^{-2(E_s-E_c)}\Theta(E_s-E_c)\ ,
\label{P_E_s}
\end{equation}
giving for the estimate of the sum
\begin{equation}
\sum_{p=1}^{P}{\rm{e}}^{\beta E_{s_p}} \sim P \int_{E_c} \!dE_s\  2\, {\rm{e}}^{-2(E_s-E_c)}\,{\rm e}^{\beta E_s}
\label{sum_estimate}
\end{equation}
For the temperatures $T<1/2$ that we are interested in here the integral diverges at the upper end, indicating that the sum is in fact dominated by its largest entries. One can estimate the largest $E_s$ among the $P$ saddles by setting the cumulative probability in the tail of~\eqref{P_E_s} to $1/P$, giving
\begin{equation}
{\rm e}^{-2(E^{\rm max}_s-E_c)} \sim 1/P
\end{equation}
The largest term in the sum in~\eqref{normalization}, including the prefactor $1/N_R$, is then of order
\begin{equation}
\frac{{\rm e}^{\beta E_s^{\rm max}}}{N_R} \sim 
\frac{1}{N_R}\left(\frac{{\rm e}^{-2 E_c}}{P}\right)^{-\beta/2} \sim N^{(\beta/2)-1}
\label{largest}
\end{equation}
where we have used that, for large cutoffs $E_c$, $N_R\sim N$  and $P\sim N{\rm e}^{-2 E_c}$. The largest element~\eqref{largest} thus diverges with $N$ for $T<1/2$, showing that the constant first term in the normalization condition~\eqref{normalization}, which comes from the remainder node, is negligible as stated in the main text. 

The fact that the sum in the normalization condition~\eqref{normalization} is dominated by its largest term also implies that the largest eigenvector entries, which occur on the deepest saddles, are of order unity. This is consistent with our assumption that the power-law tail~\eqref{px} in the distribution of $x=N\u^2$ extends up to $x$ of $O(N)$.



\section{Correlation and radial distribution function }
\label{app:corr}

\begin{figure}
  \centering
  \includegraphics[width=0.4\textwidth]{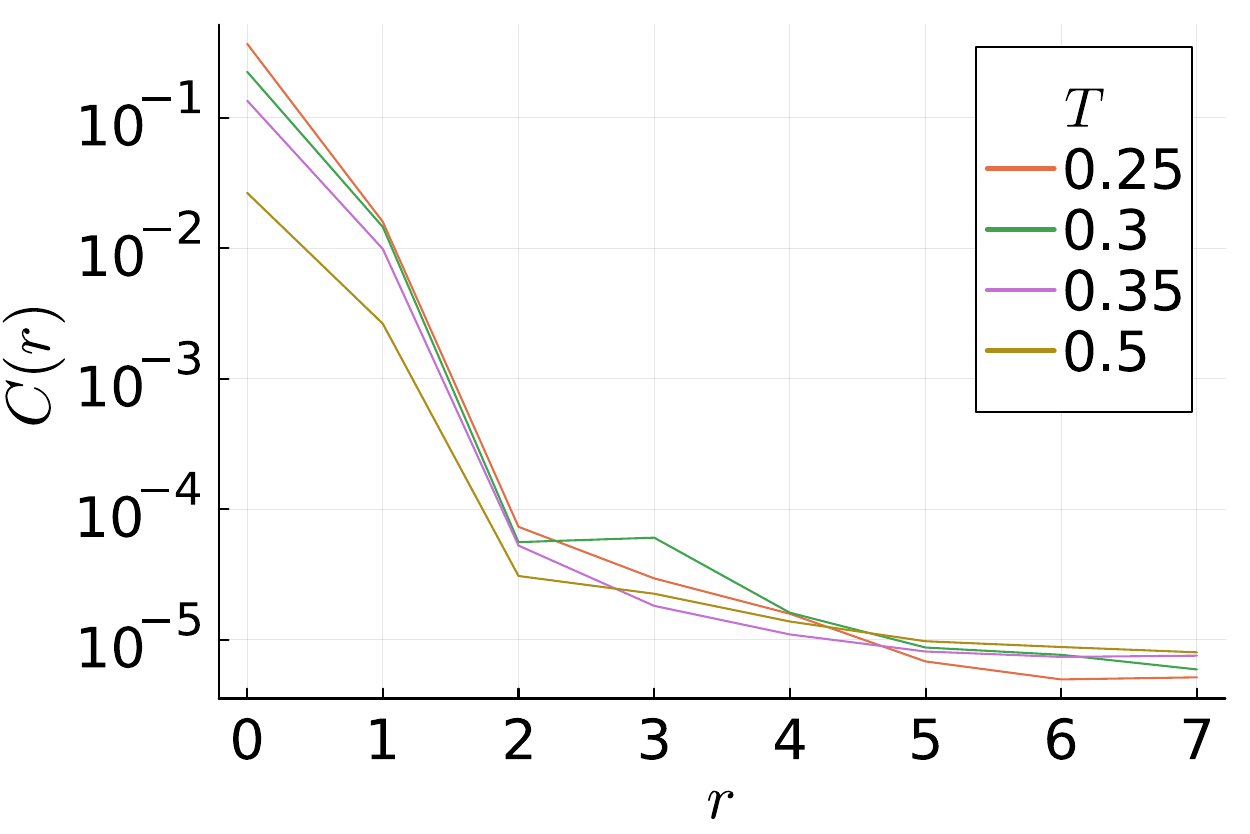}
  \caption{Correlation function (eq.~\eqref{correlation}) obtained as the average across 10 different instances of size $N=2^{17}$ at each temperature $T$, using 50 eigenvectors per instance around $\lambda = -1/3$.}
  \label{correlation_fig}
\end{figure}

\begin{figure}
  \centering
  \includegraphics[width=0.4\textwidth]{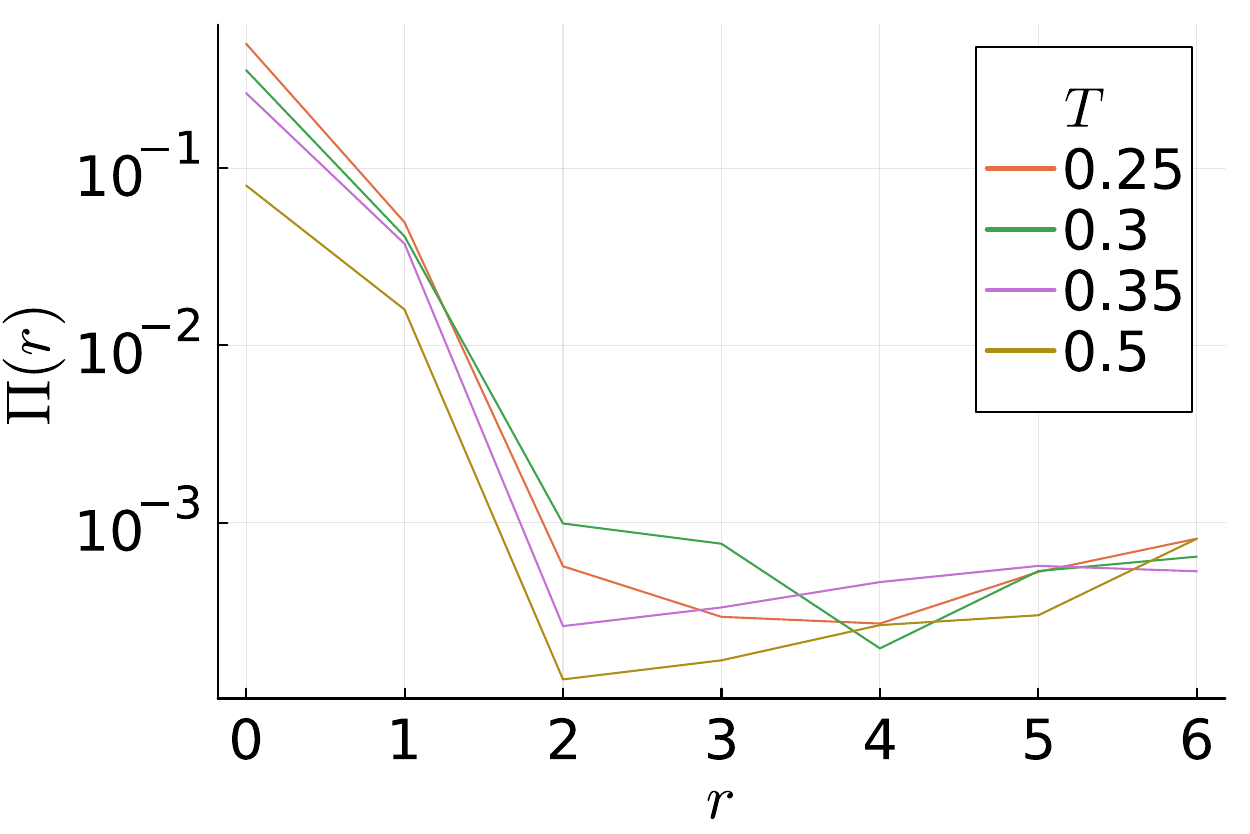}
  \caption{Radial distribution function (eq.~\eqref{radial}) obtained as the average across 10 different instances of size $N=2^{17}$
at each temperature $T$, using 50 eigenvectors per instance around $\lambda = -1/3$.}
  \label{radial_fig}
\end{figure}

We provide here some brief evidence that the localization at the spectral divergences of the sparse BM model is statistical rather than spatial in nature. In particular, we show in Figs.~\ref{correlation_fig} and~\ref{radial_fig} plots of  the correlation function and radial probability distribution, respectively. These are defined as follows. 

The distance--dependent correlation function is~\cite{tikhonov2019statistics, garcia2022critical}:
           \begin{align}
        C(r) = N \langle | \psi(0) |^2   |\psi(r) |^2 \rangle
        \label{correlation}
           \end{align}
Explicitly and for a single master operator instance, the average on the right is over all node pairs separated by a distance $d(i,j)=r$:
           \begin{align}
             \langle | \psi(0) |^2   |\psi(r) |^2 \rangle = \sum_i  \frac{1}{N_i(r)}  \sum_{j : d(i,j) = r}    | \psi_i |^2   |\psi_j |^2 
           \end{align}
           with $N_i(r)$ the number of nodes at distance $r$ from node $i$.

The radial distribution function is defined as~\cite{de2021rare, garcia2022critical}:
            \begin{align}
        \Pi(r) = \left \langle \sum_{d(i_{\max}, i) = r} |\psi_i |^2 \right \rangle
        \label{radial}
            \end{align}
            with $i_{\max}$ the site where the squared amplitude of a given wavefunction (of a given master operator instance) reaches its maximum. For the Anderson model on a random regular graph, the decay of the correlation function has been estimated analytically  as~\cite{tikhonov2019statistics}
            \begin{align}
              C(r) \sim (c-1)^{-r} \frac{\exp(-r/\xi)}{r^{3/2}}
            \end{align}
            with $\xi$ the (disorder--dependent) localization length. This in turns implies a similar decay for the radial  probability distribution, namely~\cite{garcia2022critical}
            \begin{align}
              \Pi(r) \sim \frac{\exp(-r/\xi)}{r^{3/2} }
            \end{align}

            In contrast to the Anderson model, what we see in Figs.~\ref{correlation_fig} and~\ref{radial_fig} for the sparse BM model is that there is no evidence of a nontrivial correlation length.
Instead, what the figures show is that there is a strong correlation for small lengths ($r\leq 1$) that becomes weak and largely $r$--independent for $r \geq 2$. This makes sense in view of our effective model, because the correlation may be high among node pairs (nodes separated by a distance $r = 1$) while beyond that distance there is no spatial structure that supports the localization, with the latter being based on statistical node properties rather than spatial proximity.

\section{Network--based visualization}
\label{app:net}

\begin{figure}
  \centering
  \includegraphics[width=0.4\textwidth]{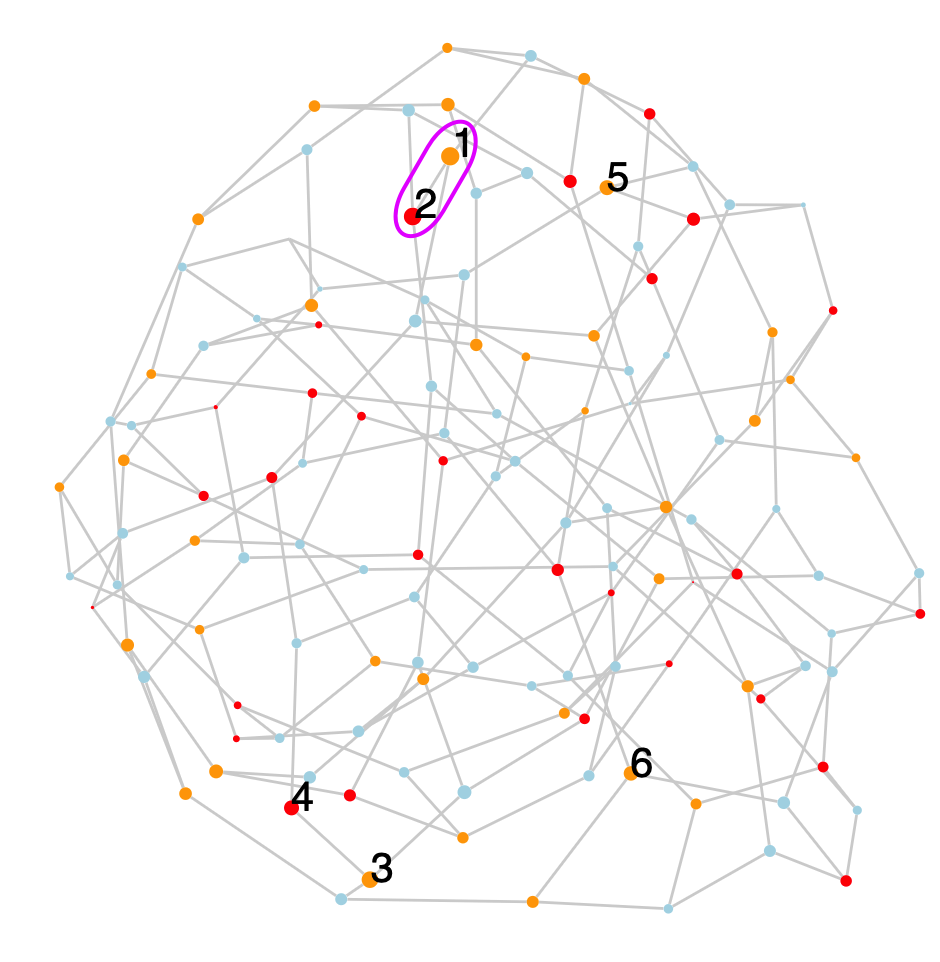}
  \caption{
Energy landscape on a random regular graph with \( N = 2^7 = 128 \) nodes. 
Local minima are coloured red, index-1 saddles orange, and local maxima or higher-index saddles blue.
The radius of each node is proportional to \( \ln x \), where \( x \) is the squared amplitude of the wavefunction with eigenvalue closest to \(-1/3\) at \( T = 0.2 \). 
The six nodes with the largest squared amplitudes are labeled \( \{1, \ldots, 6\} \) in descending order. 
The matrix of pairwise distances between these six dominant nodes is shown in equation~\eqref{distances}. 
An ellipse highlights the dominant saddle-minimum pair, which plays a key role in the low-temperature dynamics as discussed in the main text.
}
  \label{network_vis}
\end{figure}

            In Fig.~\ref{network_vis} we show a single small instance of the sparse BM model with $N = 64$ nodes, with every node painted a certain color according to its role (minimum, maximum, saddle) in the energy landscape. Additionally, the top six nodes carrying most of the ``mass'' (squared eigenvector entry) of the eigenvector with eigenvalue closest to $-1/3$ are highlighted. Computing the distance matrix for these nodes, we find
            \begin{equation}
\begin{bmatrix}
0 & 1 & 5 & 4 & 4 & 6 \\
1 & 0 & 6 & 5 & 5 & 7 \\
5 & 6 & 0 & 1 & 7 & 3 \\
4 & 5 & 1 & 0 & 6 & 4 \\
4 & 5 & 7 & 6 & 0 & 4 \\
6 & 7 & 3 & 4 & 4 & 0 \\
\end{bmatrix}
\label{distances}
\end{equation}
Interestingly, the top two entries correspond to a single saddle--minimum pair, separated by a distance of at least 4 from the other nodes with large eigenvector entries. This is a clear illustration of the statistical localization mechanism: beyond the structure of saddle--minimum pairs, nodes with large eigenfunction entries are not spatially related.

\bibliography{cavi}

\end{document}